\documentclass[12pt,english,floatfix,superscriptaddress,aps,prd,preprint,nofootinbib]{revtex4}
\usepackage{amsmath}
\usepackage{booktabs}
\usepackage{siunitx}
\usepackage{amssymb}
\usepackage{amsbsy}
\usepackage[a4paper, margin=1.8cm]{geometry}
\usepackage{amsfonts}
\usepackage{amsopn}
\usepackage{amstext}
\usepackage{graphicx}
\usepackage{amssymb}
\usepackage{amsfonts}
\usepackage{amsmath}
\usepackage{graphicx}
\usepackage[english]{babel}
\usepackage{color}
\usepackage{xcolor}
\usepackage{slashed}
\usepackage{esint}
\usepackage[dvips]{epsfig}
\usepackage[dvips]{graphicx}
\usepackage{subcaption} 
\usepackage{floatrow}
\usepackage{textcomp}
\usepackage{placeins}
\usepackage{hyperref}             
\hypersetup{
    colorlinks=true,              
    breaklinks=true,              
    citecolor=blue,               
    linkcolor=[rgb]{0,0.5,0.9},   
    urlcolor=red,                 
    filecolor=green               
}

\usepackage{hyperref}
\usepackage{slashed}

\newcommand{\ie}{\begin{equation}}
\newcommand{\fe}{\end{equation}}
 \newcommand{\bq}{\begin{equation}}
 \newcommand{\eq}{\end{equation}}
 \newcommand{\bqn}{\begin{eqnarray}}
 \newcommand{\eqn}{\end{eqnarray}}

\begin{document}

\title{Gravitational Wave Signatures from Periodic Orbits around a Non--commutative Schwarzschild Black Hole}



\author{N. Heidari}
\email{heidari.n@gmail.com}

\affiliation{Departamento de Física, Universidade Federal de Campina Grande Caixa Postal 10071, 58429-900 Campina Grande, Paraíba, Brazil.}
\affiliation{Center for Theoretical Physics, Khazar University, 41 Mehseti Street, Baku, AZ-1096, Azerbaijan.}
\affiliation{School of Physics, Damghan University, Damghan, 3671641167, Iran.}

\author{A. A. Ara\'{u}jo Filho}
\email{dilto@fisica.ufc.br}

\affiliation{Departamento de Física, Universidade Federal de Campina Grande Caixa Postal 10071, 58429-900 Campina Grande, Paraíba, Brazil.}
\affiliation{Center for Theoretical Physics, Khazar University, 41 Mehseti Street, Baku, AZ-1096, Azerbaijan.}
\affiliation{Departamento de Física, Universidade Federal da Paraíba, Caixa Postal 5008, 58051--970, João Pessoa, Paraíba,  Brazil.}


\author{Iarley P. Lobo}
\email{lobofisica@gmail.com}

\affiliation{Departamento de Física, Universidade Federal de Campina Grande Caixa Postal 10071, 58429-900 Campina Grande, Paraíba, Brazil.}
\affiliation{Departamento de Física, Universidade Federal da Paraíba, Caixa Postal 5008, 58051--970, João Pessoa, Paraíba,  Brazil.}

\author{V. B. Bezerra}
\email{valdir@fisica.ufpb.br}
\affiliation{Departamento de Física, Universidade Federal da Paraíba, Caixa Postal 5008, 58051--970, João Pessoa, Paraíba,  Brazil.}

\begin{abstract}

In this work, we investigate massive particle motion and the gravitational wave emission generated by periodic trajectories around a non--commutative \textit{Schwarzschild} black hole sourced by a Lorentzian matter distribution. We analyze the effective potential, the marginally bound orbit, and the innermost stable circular orbit, showing that non--commutative corrections shift these characteristic orbits toward smaller radii and reduce their corresponding angular momenta. The allowed region in the $(E, L)$ plane is also displaced toward lower values, favoring more tightly bound configurations. Periodic trajectories are classified through the rational parameter $q$, which relates the radial and azimuthal frequencies. For a fixed orbital topology, increasing the non--commutative parameter lowers the energy required to produce the orbit and results in more compact zoom--whirl configurations. Small deviations from the periodic energies are also shown to generate precessional drift. From the periastron advance of the S2 star around Sgr~A$^*$, we obtain the preliminary bound $\Theta/M^{2}<0.014$. Finally, using the adiabatic and numerical kludge approximations, we compute the gravitational wave polarizations and find phase shifts and an overall enhancement of the amplitude.

\end{abstract}

\maketitle

\pagebreak

\tableofcontents


\section{Introduction}

Non--commutative geometry provides two conceptually different procedures for incorporating a fundamental length scale into black hole spacetimes. The first one modifies the gravitational geometry itself, whereas the second preserves the geometric sector and introduces the deformation through the matter source. Both approaches originate from the replacement of the ordinary spacetime coordinates by operators satisfying $[\hat{x}^{\mu},\hat{x}^{\nu}] = i\Theta^{\mu\nu}$, where $\Theta^{\mu\nu}$ represents a constant antisymmetric tensor responsible for measuring the non--commutativity of spacetime.

In the geometric approach, the introduction of non--commutative corrections requires an appropriate deformation of the symmetries underlying the gravitational theory. In particular, gauge structures based on the Poincaré or de Sitter groups can be extended by employing the Seiberg--Witten map. This procedure allows the non--commutative fields to be written in terms of their commutative counterparts while preserving the gauge structure of the theory \cite{Herceg:2023pmc}. Within this construction, Chaichian \textit{et al.} obtained one of the first non--commutative extensions of the \textit{Schwarzschild} geometry \cite{chaichian2008corrections}. More recently, this gauge--gravity formulation was reconsidered and extended in Ref.~\cite{Juric:2025kjl}.

A different prescription introduces the effects of non--commutativity exclusively through the source of the gravitational field. Instead of describing the black hole mass by means of a Dirac delta distribution, Nicolini \textit{et al.} replaced the pointlike source by a regular matter distribution whose characteristic width is controlled by the non--commutative scale \cite{nicolini2006noncommutative}. Therefore, the mass is no longer concentrated at a single point but distributed over a finite region of spacetime. Two profiles commonly considered within this framework are the Gaussian distribution $\rho_{\Theta}(r) = \frac{M}{(4\pi\Theta)^{3/2}}\exp\left(-\frac{r^{2}}{4\Theta}\right)$ and the Lorentzian distribution $\rho_{\Theta}(r) = \frac{M\sqrt{\Theta}}{\pi^{3/2}\left(r^{2}+\pi\Theta\right)^{2}}$. 
In both cases, the standard point--mass description is recovered in the limit in which the non--commutative smearing scale becomes negligible.

These different implementations have motivated several investigations concerning the possible consequences of non--commutativity in black hole physics. In particular, modifications have been found in the evaporation process and in the corresponding lifetime of black holes \cite{myung2007thermodynamics,23araujo2023thermodynamics}. Corrections have also been reported for the entropy, heat capacity, and other thermodynamic quantities \cite{nozari2007thermodynamics,banerjee2008noncommutative,nozari2006reissner,sharif2011thermodynamics,Heidari:2025iiv,araujo20sss25non,Heidari_2026}. Moreover, the influence of the non--commutative parameter on the quasinormal spectrum and on the perturbative response of black hole spacetimes has been widely investigated \cite{zhao2023quasinormal,anacleto2023absorption,Anacleto:2019tdj,heidari2024exploring,mann2011cosmological,campos2022quasinormal,anacleto2021quasinormal,asdasd2,karimabadi2020non,lopez2006towards,modesto2010charged,nicolini2009noncommutative,AraujoFilho:2024rss,Filho:2024zxx}. More recently, non--commutative corrections have also been considered in the construction, analysis of thin--shell gravastar configurations \cite{Anacleto:2026hkb} and neutrino physics \cite{AraujoFilho:2024mvz,AraujoFilho:2025rzh}.

Periodic geodesics have become an important tool for investigating the strong--field region of black hole spacetimes. In contrast to generic bound trajectories, periodic orbits can be classified by the relation between the radial and azimuthal frequencies, providing a systematic description of their orbital structure. This classification includes highly relativistic zoom--whirl configurations, in which the orbiting body alternates between large radial excursions and several revolutions close to the black hole. Since the properties of these trajectories depend directly on the spacetime geometry, their analysis can reveal how modifications of the gravitational background affect the topology of the motion, the orbital precession, and the behavior of particles near the central object \cite{Gong:2025mne,Kumar:2024our,Zhang:2025wni,Junior:2024tmi,Lu:2025cxx,Wang:2025hla}.

The interest in periodic motion is not restricted to the geodesic structure. The orbital frequencies and the number of whirls performed near the black hole influence the phase evolution, harmonic content, and amplitude of the gravitational radiation emitted by the orbiting body. Therefore, changes in the underlying geometry may be transferred to the corresponding gravitational wave signal. This connection makes periodic trajectories particularly useful for identifying possible deviations from general relativity and for searching for signatures of additional gravitational fields, quantum corrections, or other effects beyond the standard black hole description \cite{Addazi:2021xuf,Chen:2025aqh,Ahmed:2025shr,Shi:2026zxx,Liu:2025swi,Li:2025sfe}.

Motivated by these aspects, periodic orbits have been investigated in several black hole backgrounds. Among them, one may mention static and rotating regular black holes \cite{Gong:2025mne,Kumar:2024our}, the Dastan--Destounis--Suvorov--Kokkotas geometry \cite{Hua:2026kvw}, the Schwarzschild--Bertotti--Robinson spacetime \cite{Xamidov:2026kqs}, and black holes obtained within effective field theory scenarios \cite{Alloqulov:2025dqi}. Related analyses have also been performed for dyonic ModMax black holes \cite{Alloqulov:2025dqi}, the $\gamma$ metric \cite{Zhang:2025wni}, and different quantum--corrected configurations \cite{Chen:2025aqh,Ahmed:2025shr,Mohammadi:2026xvg}.

Periodic motion has further been explored in gravitational models containing additional fields or symmetry--breaking sectors. These investigations include Einstein--Æther black holes \cite{Lu:2025cxx}, black holes arising in bumblebee gravity \cite{Shi:2026zxx,Liu:2025swi}, and solutions supported by the Kalb--Ramond field \cite{Junior:2024tmi,Xia:2025yzg}. Other examples involve charged black holes endowed with scalar hair \cite{Deng:2025wzz}, renormalization--group--improved Kerr geometries \cite{Li:2025sfe,kumar2024exploring}, black holes without a Cauchy horizon \cite{Wang:2025hla}, and solutions obtained in the context of asymptotically safe gravity \cite{Kumar:2026hfx}. Besides modifications of the gravitational theory itself, the surrounding environment may also change the orbital dynamics and the associated gravitational radiation. For this reason, recent works have considered periodic trajectories around black holes immersed in dark matter distributions \cite{Haroon:2025rzx,Hassanabadi:2026wku,Li:2025eln,Alloqulov:2025ucf,heidari2026gravitational} and non--commutative inspired black holes surrounded by quintessence \cite{Ahmed:2025azu}.

To the best of our knowledge, the gravitational wave signals generated by periodic orbits around a non--commutative \textit{Schwarzschild} black hole sourced by a Lorentzian matter distribution have not yet been investigated. In this manner, the present work aims to fill this gap by establishing a connection between the non--commutative deformation, the strong--field orbital dynamics, and the corresponding gravitational radiation. Initially, we analyze the timelike geodesics and determine the conditions associated with bound motion, including the marginally bound orbit and the innermost stable circular orbit. Subsequently, the periodic trajectories are classified through the rational parameter $q$, and small deviations from exact periodicity are considered to examine the resulting orbital precession. We also use the periastron advance of the S2 star around Sgr~A$^*$ to obtain a preliminary constraint on the non--commutative parameter. Finally, within the adiabatic and numerical kludge approximations, we calculate the gravitational wave polarizations generated by representative periodic orbits in the extreme mass--ratio regime and investigate how the non--commutative corrections are encoded in the corresponding waveforms.

\section{Non--commutative Schwarzschild Black Hole}
To incorporate non--commutative effects, we adopt the Lorentzian smeared matter distribution, whose energy density is given by \cite{nozari2007thermodynamics}
\begin{equation}
\rho_{\Theta}(r)=\frac{M\sqrt{\Theta}}
{\pi^{3/2}\left(r^{2}+\pi\Theta\right)^{2}}.
\end{equation}
In this framework, the gravitational source is no longer localized at a point but is distributed over a region of characteristic width $\sqrt{\Theta}$, reflecting the intrinsic nonlocality of the underlying spacetime. The corresponding mass function is obtained by integrating the density over a spherical volume
\begin{equation}
M(r,\Theta)=4\pi\int_{0}^{r}\rho_{\Theta}(r')\,r'^2\,dr',
\end{equation}
which yields
\begin{equation}
M(r,\Theta)=\frac{2M}{\pi}
\left[
\tan^{-1}\!\left(\frac{r}{\sqrt{\pi\Theta}}\right)
-\frac{r\sqrt{\pi\Theta}}{r^{2}+\pi\Theta}
\right].
\end{equation}

For $r^{2}\gg \pi\Theta$, the mass function can be expanded as
\begin{equation}
M(r,\Theta)\simeq
M-\frac{4M\sqrt{\Theta}}{\sqrt{\pi}\,r},
\end{equation}
indicating that non--commutative corrections are relevant only at short distances and disappear asymptotically. Substituting this expression into the \textit{Schwarzschild} metric leads to the non--commutative line element with metric function
\begin{equation}\label{fNC}
f(r,\Theta)=1-\frac{2M}{r}
+\frac{8M\sqrt{\Theta}}{\sqrt{\pi}\,r^{2}}.
\end{equation}
The additional $r^{-2}$ term represents the leading-order contribution of spacetime non--commutativity, while the classical \textit{Schwarzschild} geometry is recovered in the commutative limit $\Theta\to0$. Hereafter, ``\textit{NC}'' will be used as an abbreviation for ``non--commutative'', and the resulting geometry will be referred to as the ``\textit{NC Schwarzschild}" black hole. Moreover, for convenience, the \textit{NC} parameter is expressed in terms of the dimensionless ratio $\Theta/M^2$ throughout this work.


\section{Timelike geodesics}

The motion of a neutral test particle in a static and spherically symmetric background spacetime can be formulated within the Hamiltonian formalism. The corresponding Hamiltonian is given by
\begin{equation}
\mathcal{H} = \frac{1}{2}     g^{\mu\nu}p_{\mu}p_{\nu},
\end{equation}
where $p_{\mu}$ denotes the canonical four--momentum associated with the particle trajectory. The overdot represents differentiation with respect to the affine parameter $\tau$. For timelike geodesics, the Hamiltonian satisfies the normalization condition $  2\mathcal{H}=-1$, the canonical momenta conjugate to the coordinates $x^{\mu}$ are defined through $p_{\mu}=g_{\mu\nu}\dot{x}^{\nu}$, and the quantities associated with the Killing vectors $\partial_{t}$ and $\partial_{\phi}$ are conserved. These constants of motion, for the spacetime characterized by $\mathrm{d}s^2=-f(r)\mathrm{d}t^2+f(r)^{-1}\mathrm{d}r^2+r^2\mathrm{d}\Omega^2$ are identified, respectively, as the particle energy $E$ and the angular momentum $L$ in the following form
\begin{align}
p_{t} &=-f(r)\dot{t}=-E, \\
p_{\phi} &= r^{2}\sin^{2}\theta\,\dot{\phi}=L.
\end{align}

The equations of motion follow from Hamilton's equations as
\begin{equation}
\dot{x}^{\mu}  = \frac{\partial \mathcal{H}}{\partial p_{\mu}}, \qquad \dot{p}_{\mu}     = -\frac{\partial \mathcal{H}}{\partial x^{\mu}},
\end{equation}
which leads to
\begin{align}\label{eq:tdot}
\dot{t} &= \frac{E}{f(r)}, \\ \label{eq:phidot}
\dot{\phi} &= \frac{L}{r^{2}\sin^{2}\theta}.
\end{align}

Substituting the conserved quantities into the Hamiltonian constraint yields the radial equation of motion
\begin{equation}
\dot{r}^{2} + f(r) \left(1+ r^{2}\dot{\theta}^{2} + \frac{L^{2}}{r^{2}\sin^{2}\theta} \right) = E^{2}.
\end{equation}

Without loss of generality, the particle motion can be restricted to the equatorial plane by setting
$\theta=\frac{\pi}{2}$ and $  \dot{\theta}=0$. The radial equation then simplifies to
\begin{equation}\label{eq:rdot}
\dot{r}^{2}+V_{\mathrm{eff}}=E^{2},
\end{equation}
where the effective potential takes the form
\begin{equation}\label{eq:veff1}
V_{\mathrm{eff}} = f(r) \left( 1+\frac{L^{2}}{r^{2}}\right).
\end{equation}
If we substitute the metric function $f(r)$ with the \textit{NC} lapse function $f(r,\Theta)$ in Eq.~\eqref{fNC}, the effective potential takes the following form
\begin{equation}\label{eq:veff2}
V_{\mathrm{eff}} =\left(1-\frac{2 M}{r}+\frac{8 \sqrt{\Theta } M}{\sqrt{\pi } r^2}\right)\left( 1+\frac{L^{2}}{r^{2}} \right)
\end{equation}

\begin{figure}[ht!]
    \centering
    \includegraphics[width=80mm]{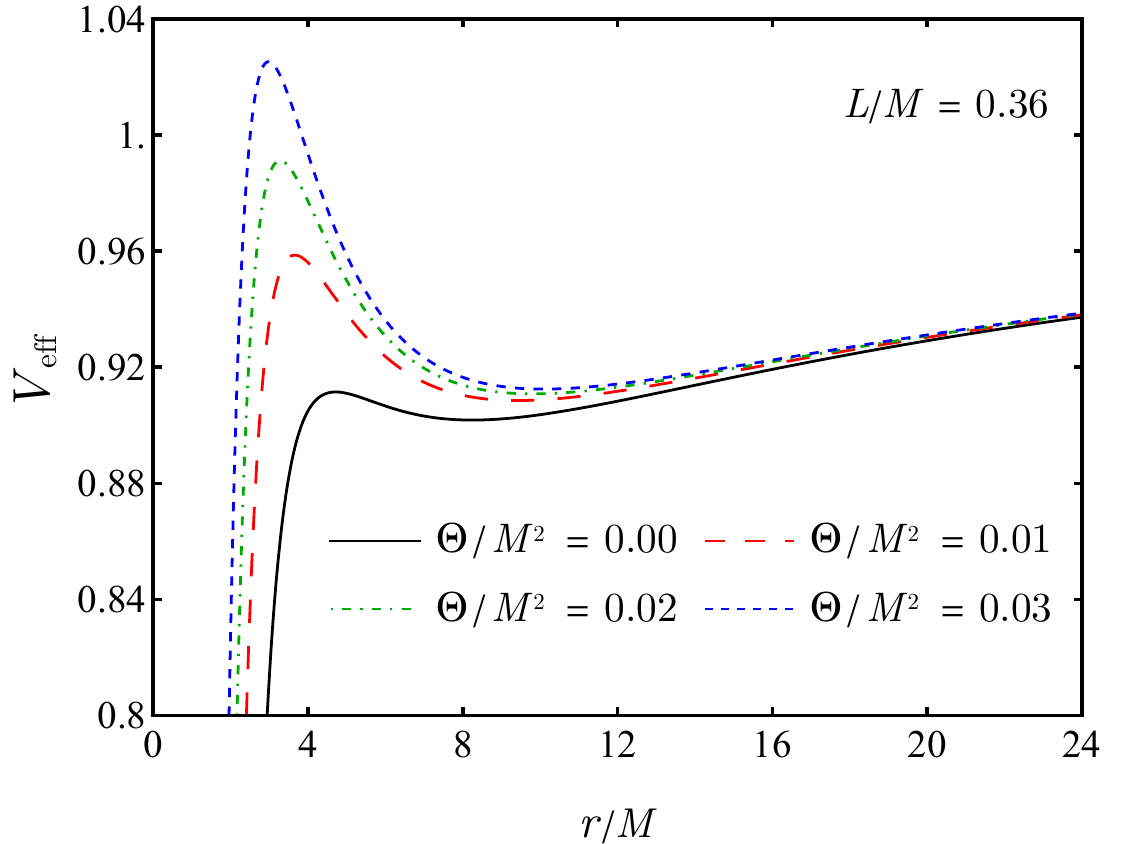}
    \includegraphics[width=80mm]{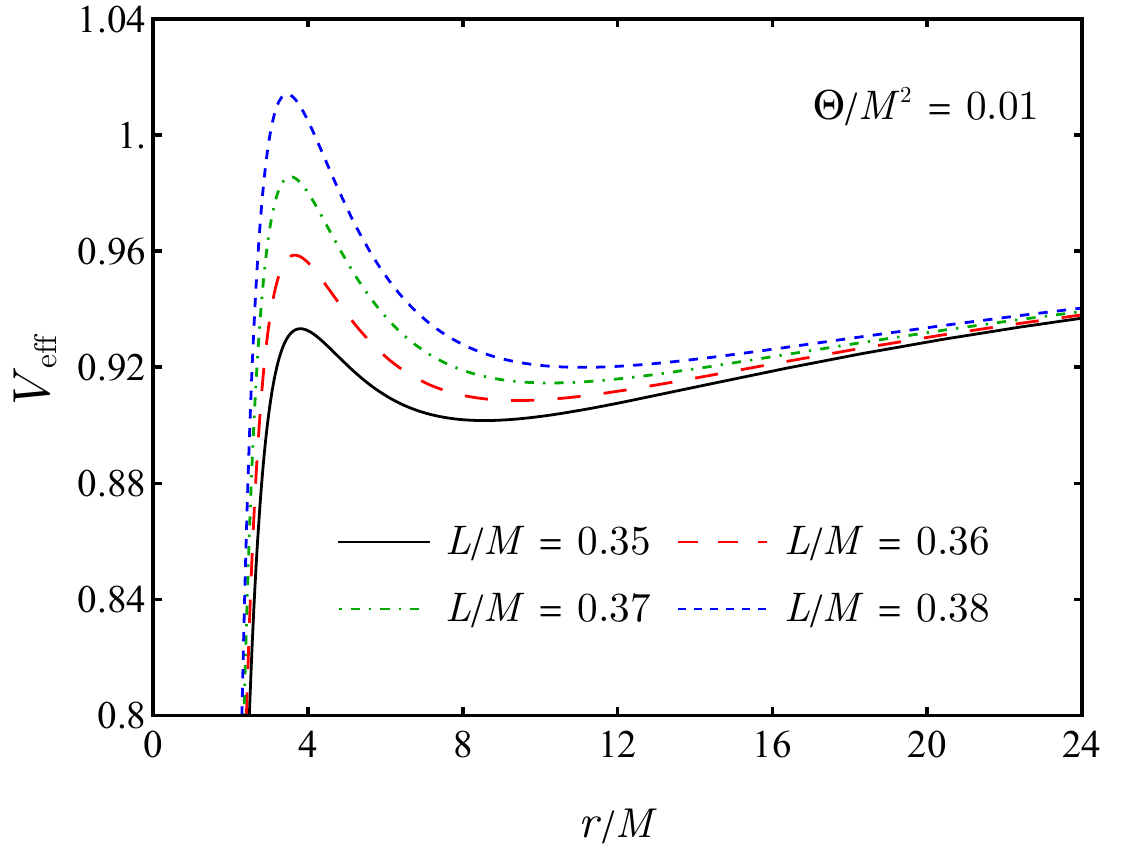}
    \protect\caption{Effective potential as a function of the radial coordinate for varying $\Theta/M^2$ at fixed angular momentum (left panel) and varying angular momentum at fixed $\Theta/M^2$ (right panel).}
    \label{fig:Veff}
\end{figure}

The radial motion and subsequently the structure of bound or unbound timelike geodesics is determined by the behavior of the effective potential. Since the spacetime is asymptotically flat, the effective potential approaches unity in the large--distance limit,
\begin{equation}
V_{\rm eff}(r)\xrightarrow{r\to\infty}1.
\end{equation}
In this manner, the conserved energy $E=1$ represents the critical threshold separating bound and unbound trajectories. From the radial equation of motion in Eq.~\eqref{eq:rdot}, particles with $E>1$ retain nonvanishing radial velocity at spatial infinity and therefore escape the gravitational field, whereas bound motion is restricted to trajectories satisfying $E\leq1$. The physically allowed domain of stable circular motion is bounded by two characteristic orbits, namely the marginally bound orbit (MBO) and the innermost stable circular orbit (ISCO). The corresponding conserved quantities satisfy
\begin{equation}\label{eq:ELconstraint}
E_{\rm ISCO} \leq  E  \leq  E_{\rm MBO}=1, \qquad  L_{\rm ISCO}\leq L.
\end{equation}
The upper bound $E_{\rm MBO}=1$ corresponds to the limiting circular orbit separating bound motion from escape trajectories, while $E_{\rm ISCO}$ determines the onset of radial instability. Similarly, circular orbits with angular momentum below $L_{\rm ISCO}$ cannot remain stable and eventually plunge into the black hole.

The MBO is determined from the conditions
\begin{equation}
V_{\rm eff}(r)=1, \qquad \frac{\mathrm{d}V_{\rm eff}}{\mathrm{d}r}=0,
\end{equation}
which yield the corresponding radius and angular momentum in the following form
\begin{align}
r_{\rm MBO} &= \frac{ 2f(r_{\rm MBO}) \left[ 1-f(r_{\rm MBO}) \right] } { f'(r_{\rm MBO}) }, \label{eq:rMBO} \\ L_{\rm MBO} &= r_{\rm MBO} \sqrt{ \frac{ 1-f(r_{\rm MBO}) } { f(r_{\rm MBO}) } }.
    \label{eq:LMBO}
\end{align}

The ISCO is obtained from the circular orbit conditions, together with the requirement that the effective potential develops an inflection point which satisfies the following conditions
\begin{equation}
\dot r =0, \qquad \frac{\mathrm{d}V_{\rm eff}}{\mathrm{d}r}=0, \qquad \frac{\mathrm{d}^{2}V_{\rm eff}}{\mathrm{d}r^{2}}=0.
\end{equation}
Substituting the effective potential from Eq.~\eqref{eq:veff1} into the above relations leads to the following equation governing the ISCO radius
\begin{equation}
r_{\rm ISCO}f(r_{\rm ISCO})f''(r_{\rm ISCO}) - 2r_{\rm ISCO}f'(r_{\rm ISCO})^{2} + 3f(r_{\rm ISCO})f'(r_{\rm ISCO})=0,
\end{equation}
and the associated angular momentum and conserved energy are given by
\begin{align}
L_{\rm ISCO} &= \sqrt{ \frac{ r_{\rm ISCO}^{3}f'(r_{\rm ISCO}) } { 2f(r_{\rm ISCO}) - r_{\rm ISCO}f'(r_{\rm ISCO}) } }, \\ E_{\rm ISCO} &= \sqrt{ \frac{ 2f(r_{\rm ISCO})^{2} } { 2f(r_{\rm ISCO}) - r_{\rm ISCO}f'(r_{\rm ISCO}) } }.
\end{align}

\begin{figure}[t]
    \centering
    \includegraphics[scale=0.4]{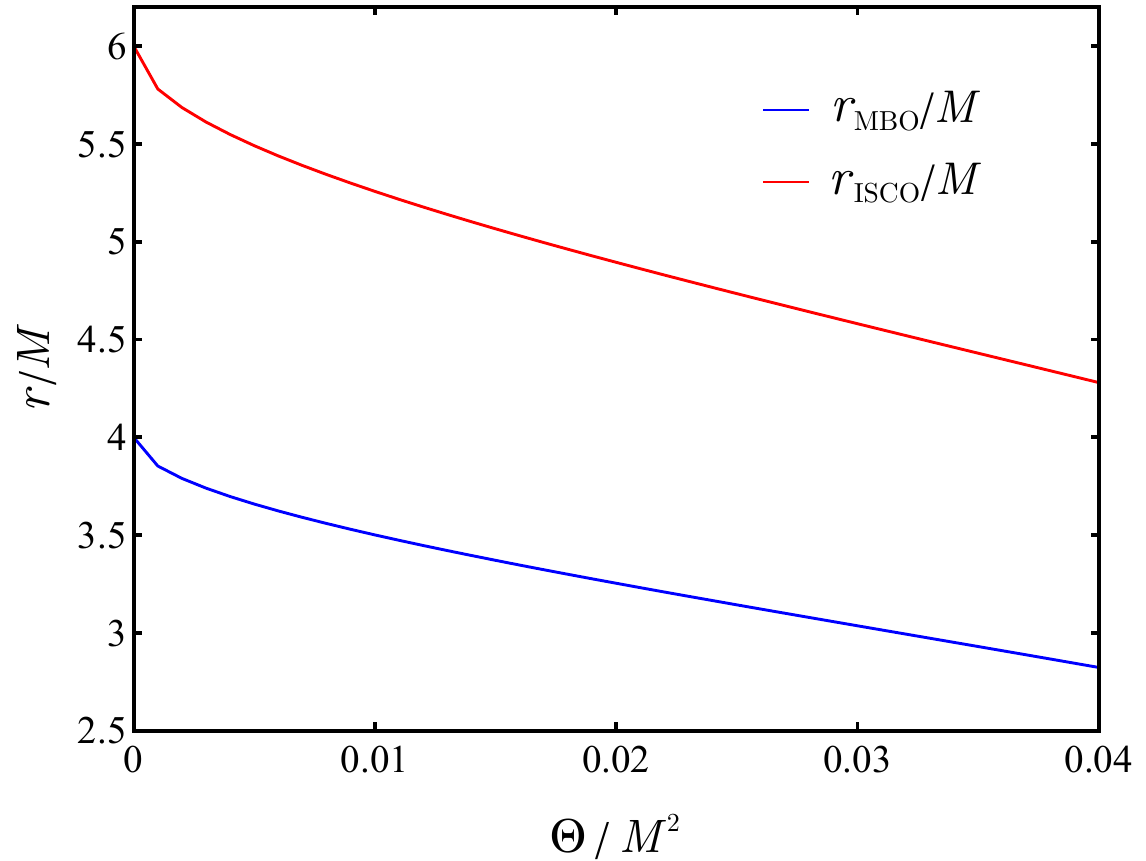}
    \includegraphics[scale=0.4]{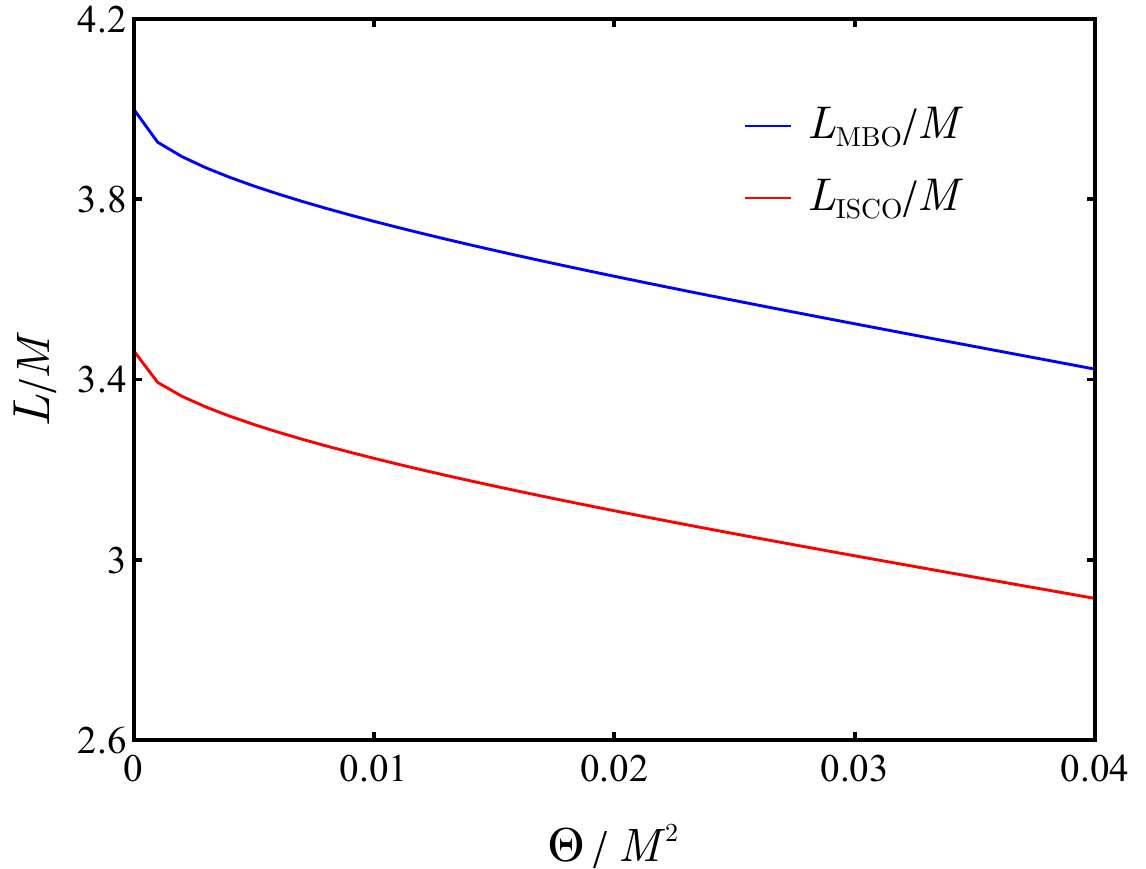}
    \caption{
Variation of the scaled MBO and ISCO radius and angular momentum with the \textit{NC} parameter $\Theta/M^{2}$.}
    \label{fig:MBOISCO}
\end{figure}

Upon substituting \textit{NC} lapse function, the above equations become analytically intractable. We utilize the numerical method and the resulting behavior of the characteristic orbital parameters is presented in Fig.~\ref{fig:MBOISCO}. The left panel shows the variation of the $r_{\rm MBO}/M$ and $r_{\rm ISCO}/M$ with the \textit{NC} parameter $\Theta/M^{2}$, while the right panel illustrates the corresponding scaled angular momenta $L_{\rm MBO}/M$ and $L_{\rm ISCO}/M$. We find that increasing the \textit{NC} parameter shifts both characteristic orbits toward smaller radius and simultaneously decreases the associated angular momentum. This behavior indicates that \textit{NC} effects allow bound circular trajectories to exist closer to the event horizon compared with the \textit{Schwarzschild} spacetime. 

\begin{figure}[t]
    \centering
    \includegraphics[width=0.6\textwidth]{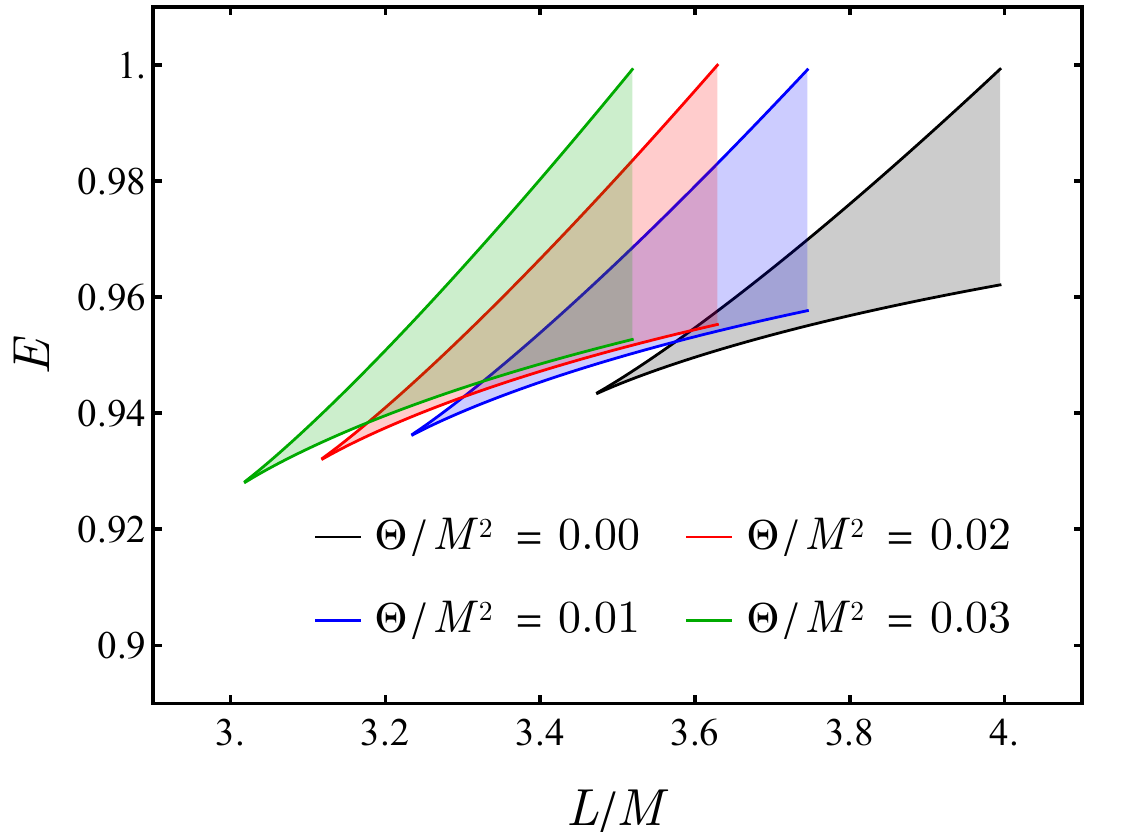}
    \caption{
    Allowed parameter space for bound timelike motion in the $(E,L)$--plane for several values of the \textit{NC} parameter $\Theta/M^{2}$. }
    \label{fig:EL}
\end{figure}

On the other hand, Eq.~\eqref{eq:ELconstraint} specifies the parameter space in the $(E,L)$--plane where bound timelike motion is possible. The limiting curves of this region are determined by MBO and ISCO orbits, as shown in Fig.~\ref{fig:EL}.  
The comparison between the \textit{Schwarzschild} and \textit{NC Schwarzschild} geometries highlights a clear impact of non--commutativity on the orbital structure. In particular, increasing the \textit{NC} parameter shifts the allowed domain toward smaller values of angular momentum. At the same time, the minimum energy associated with bound trajectories decreases, indicating that \textit{NC} effects favor more tightly bound orbital configurations, which will be explored in the next section.

\section{Periodic orbits}

Periodic orbits provide a useful framework for characterizing bound particle motion in strong gravitational fields. Although generic bound trajectories are not generally closed, they can often be approximated by nearby periodic configurations. Consequently, the study of periodic orbits offers valuable insight into the orbital structure of the underlying spacetime and has been widely employed in investigations of relativistic dynamics around compact objects.

To analyze the periodic structure of bound timelike motion, we employ the rational number classification introduced in Ref.~\cite{levin2008periodic}. In this approach, each periodic trajectory is associated with a triplet of integers $(z,w,v)$ that encodes its geometric properties. Here, ``$z$'' denotes the number of radial oscillations required for the orbit to close, ``$w$'' represents the number of whirls accumulated near the central object, and ``$v$'' determines the ordering of successive vertices.

The classification is based on the ratio between the azimuthal and radial frequencies. Defining
\begin{equation}
q \equiv \frac{\omega_{\phi}}{\omega_{r}}-1,
\end{equation}
periodic orbits correspond to rational values of $q$, which can be written as
\begin{equation}
q=w+\frac{v}{z}.
\end{equation}
The quantity $q$ therefore provides a convenient characterization of the orbital topology.

The frequency ratio is directly related to the azimuthal angle accumulated during one complete radial cycle $\frac{\omega_{\phi}}{\omega_{r}}=
\frac{\Delta\phi}{2\pi}$, where $\Delta\phi$ denotes the total angular advance between two successive passages through the periapsis. Therefore
\begin{equation}
q=\frac{\Delta\phi}{2\pi}-1.
\end{equation}
Using the equations of motion, one obtains
\begin{equation}
q=\textbf{} \frac{1}{2\pi} \int_{r_1}^{r_2} \frac{\dot{\phi}}{\dot r}\mathrm{d}r -1=\frac{1}{\pi} \int_{r_1}^{r_2} \frac{L}{r^{2}\sqrt{E^{2}-V_{\rm eff}}}\mathrm{d}r -1,
\end{equation}
This relation explicitly connects the periodic orbital structure to the conserved quantities of the particle $E$, $L$, and the properties of the underlying spacetime. Thus, variations in either the orbital constants or the gravitational background modify the value of $q$.

\begin{figure}[ht!]
\centering
\includegraphics[width=85mm]{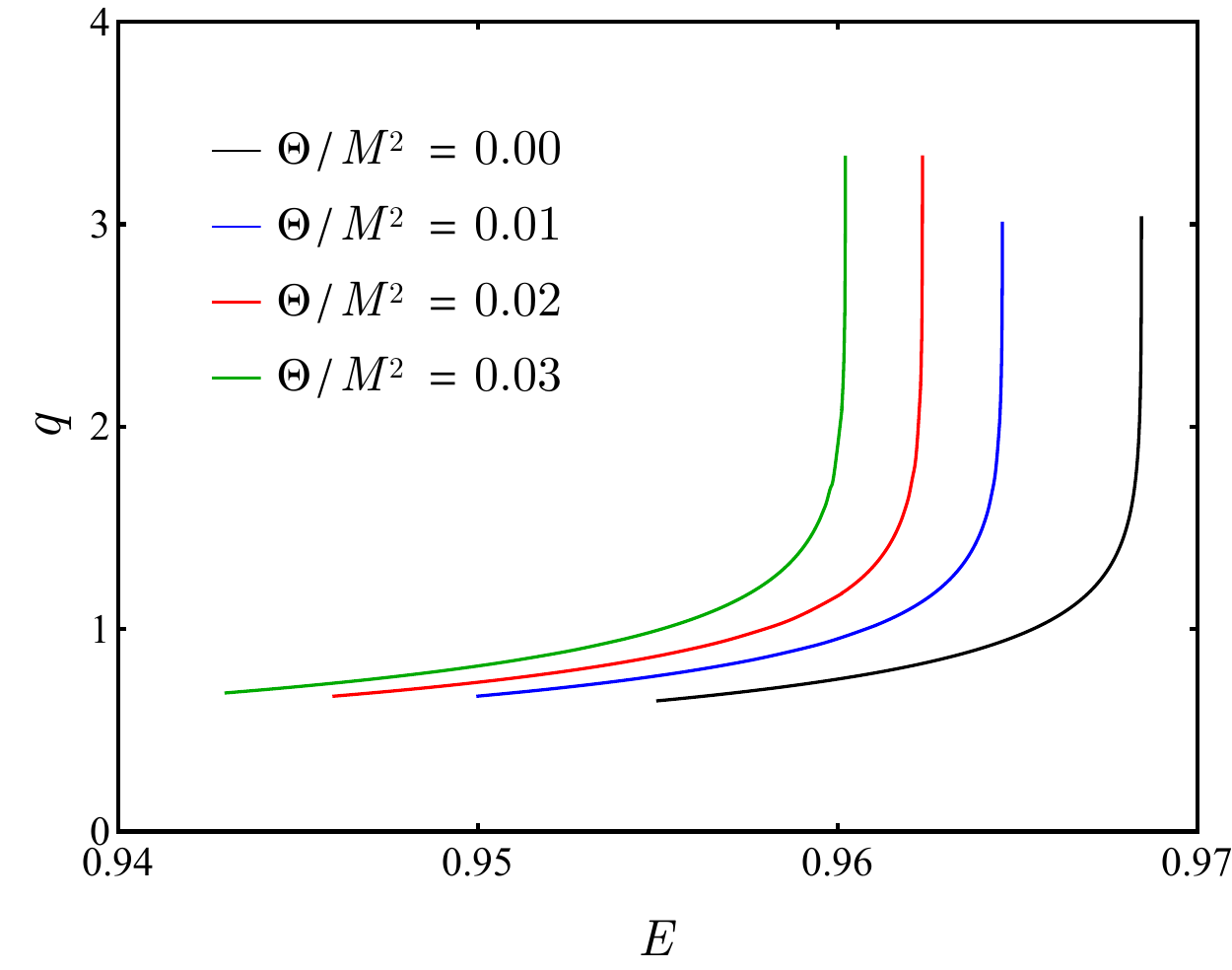}
\caption{Rational number $q$ as a function of the orbital energy $E$ for periodic bound orbits in the \textit{NC Schwarzschild} black hole spacetime. The angular momentum is fixed at $L_{\mathrm{av}}$, and the curves correspond to selected values of the \textit{NC} parameter $\Theta/M^2=0.00,0.01,0.02$ and $0.03$.}
\label{fig:qE}
\end{figure}

For bound motion, the angular momentum is confined to a finite domain determined by the ISCO and MBO values. We describe this domain by introducing a normalized parameter $\epsilon$ according to
\begin{equation}
L(\epsilon)=L_{\mathrm{ISCO}}+\epsilon
\left(L_{\mathrm{MBO}}-L_{\mathrm{ISCO}}\right),
\end{equation}
where $0\leq \epsilon \leq 1$. Then, $\epsilon=0$ selects the ISCO angular
momentum, while $\epsilon=1$ gives the MBO value. In our calculations, for simplicity, we fix the angular momentum to the central value of this interval
\begin{equation}
L_{\mathrm{av}}=\frac{L_{\mathrm{ISCO}}+L_{\mathrm{MBO}}}{2},
\end{equation}
corresponding to $\epsilon=1/2$. The role of the orbital energy $E$ in determining the rational number $q$ can be examined for each spacetime configuration, as
illustrated in Fig.~\ref{fig:qE}. In all cases, the rational number $q$ grows monotonically with the orbital energy $E$. This growth is relatively mild at lower energies, but becomes increasingly rapid as the upper part of the allowed energy range is approached. A comparison between different values of the \textit{NC} parameter $\Theta/M^2$, shows that the $q(E)$ profiles are displaced toward smaller energies as $\Theta/M^2$ increases. 

To further quantify this effect, a numerical analysis of the energy associated with selected periodic orbits is performed. Keeping the angular momentum fixed at $L_{\mathrm{av}}$, we consider several orbital configurations specified by the triplet $(z,w,v)$. The corresponding values of the energy are summarized in Table~\ref{tab:qE}. For each configuration, increasing the \textit{NC} parameter leads to a systematic reduction in the energy required to realize the same periodic orbit, in agreement with the trend observed in Fig.~\ref{fig:qE}. In other words, stronger \textit{NC} corrections allow periodic orbits with a given rational number $q$ to occur at lower orbital energies. Moreover, the associated averaged angular momentum $L_{\mathrm{av}}$ increases by higher valuesthe  of \textit{NC} parameter.

\begin{figure}[htbp]
    \centering
    \setlength{\tabcolsep}{4pt}
    
    \begin{tabular}{cccc}
        \includegraphics[width=0.23\textwidth]{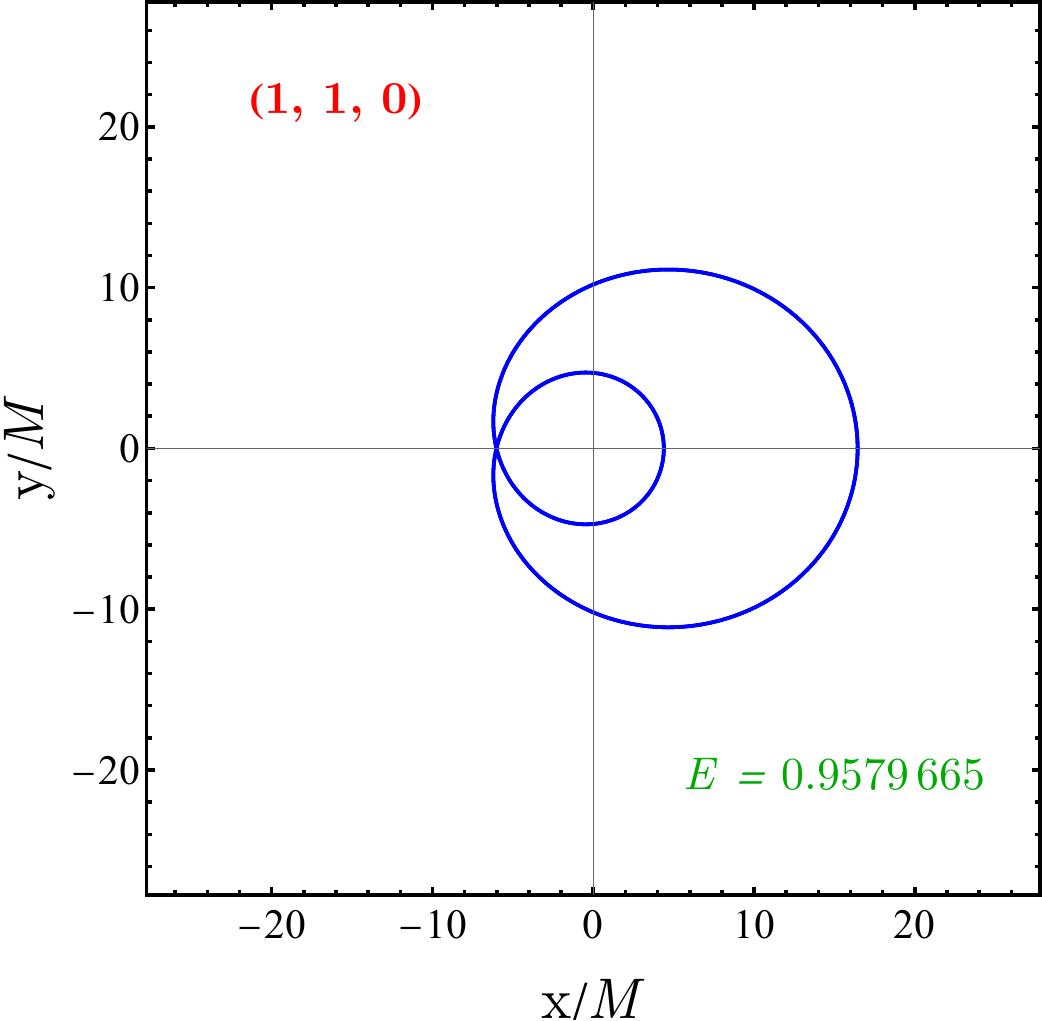} &
        \includegraphics[width=0.23\textwidth]{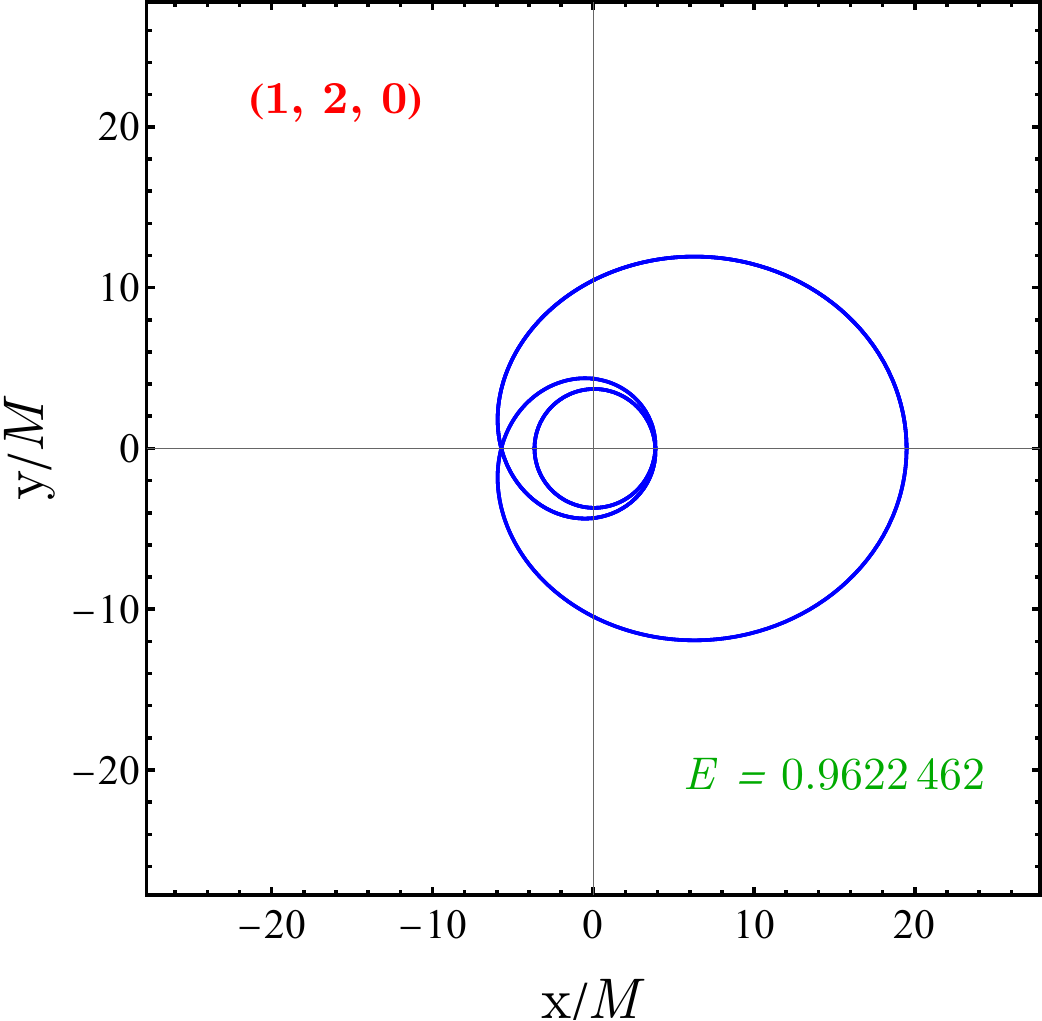} &
        \includegraphics[width=0.23\textwidth]{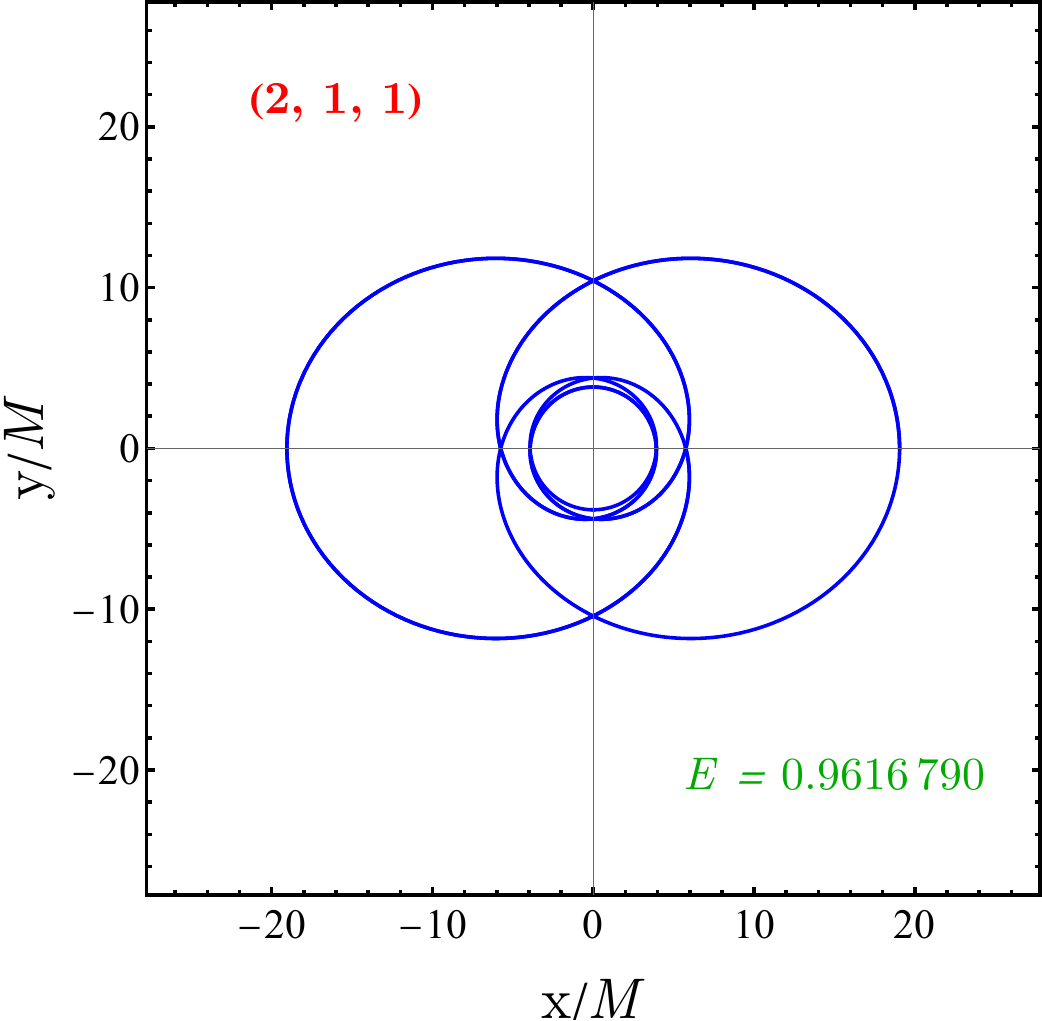} &
        \includegraphics[width=0.23\textwidth]{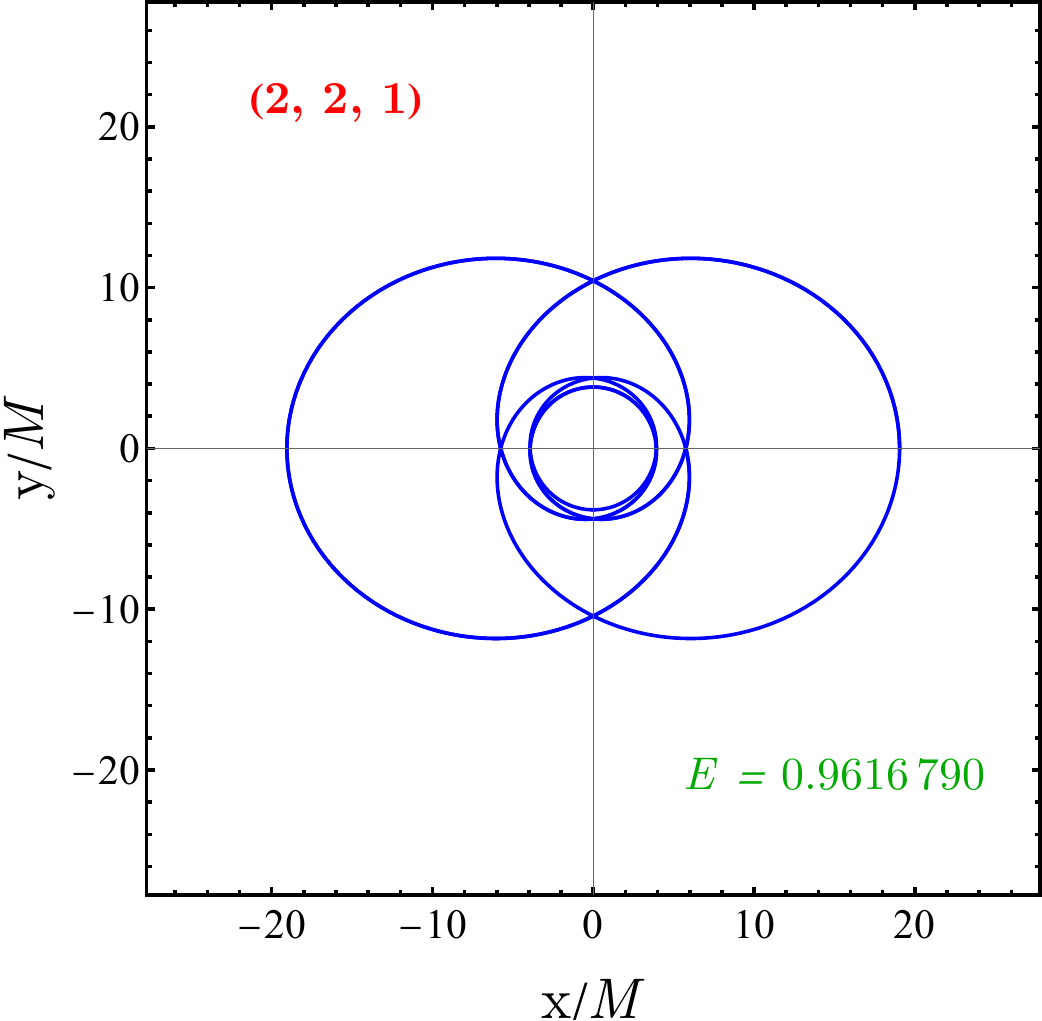} \\

        \includegraphics[width=0.23\textwidth]{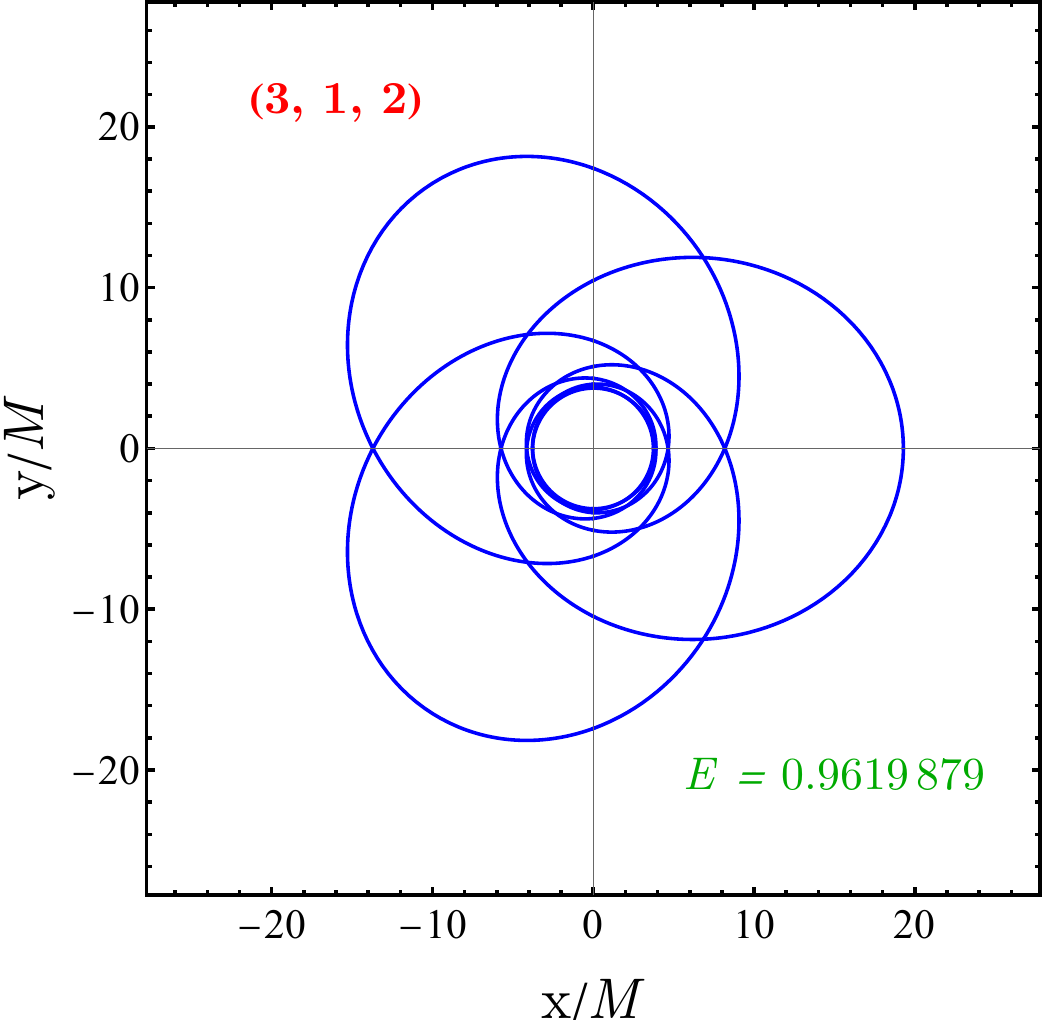} &
        \includegraphics[width=0.23\textwidth]{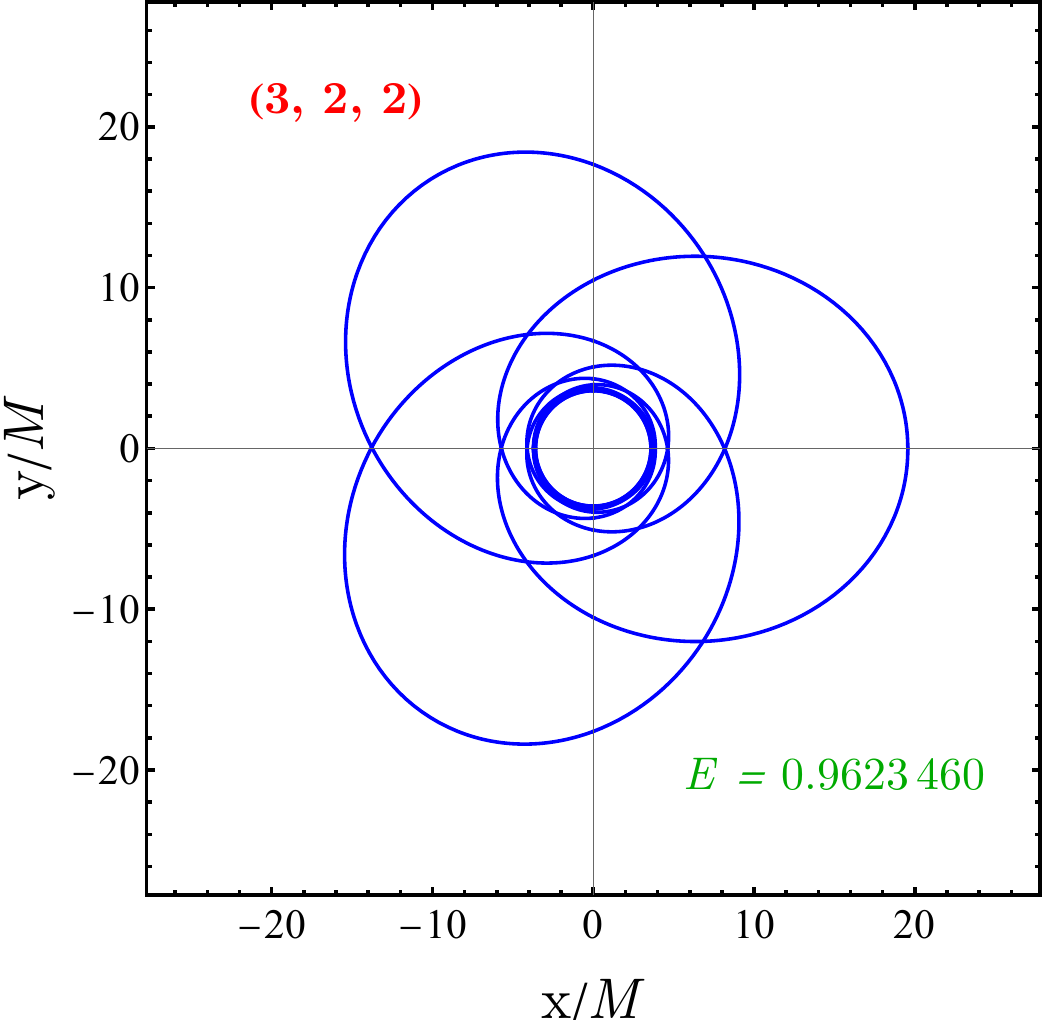} &
        \includegraphics[width=0.23\textwidth]{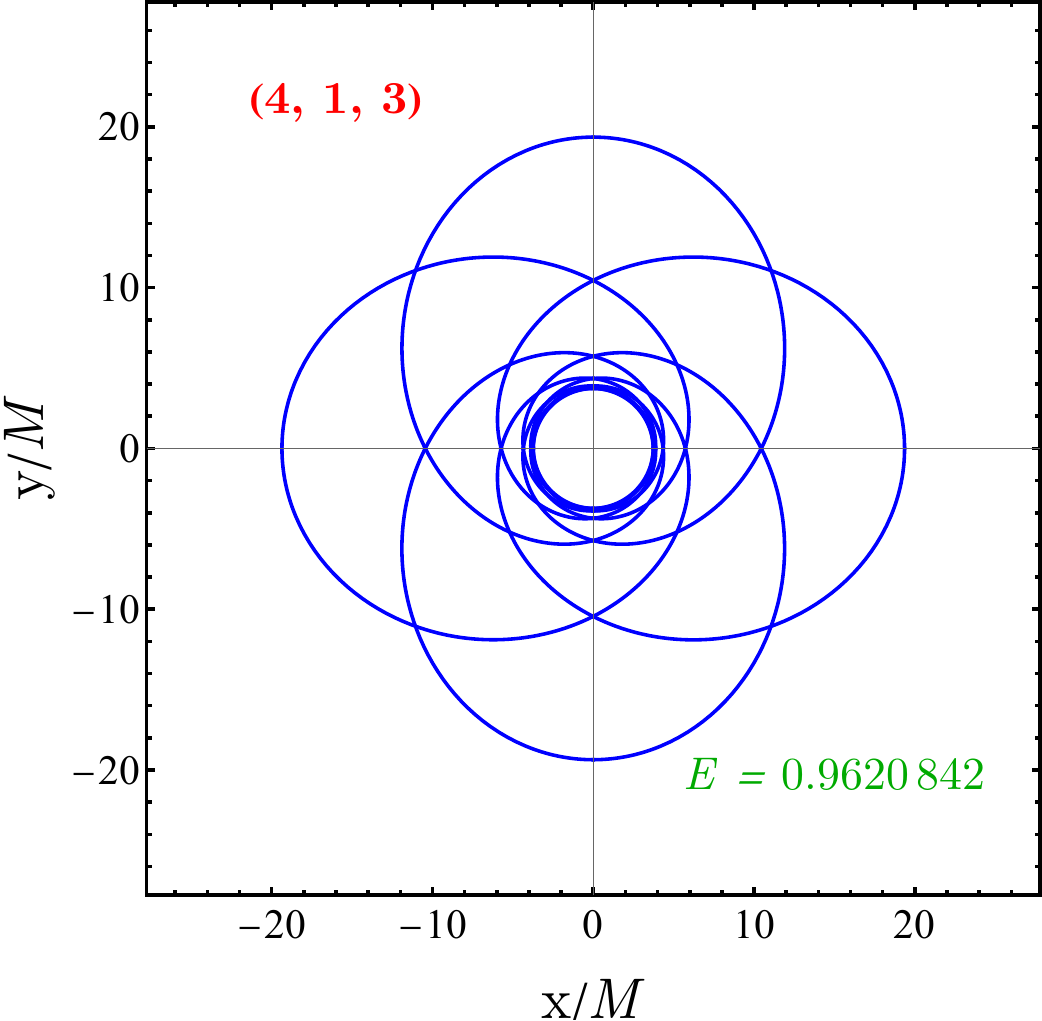} &
        \includegraphics[width=0.23\textwidth]{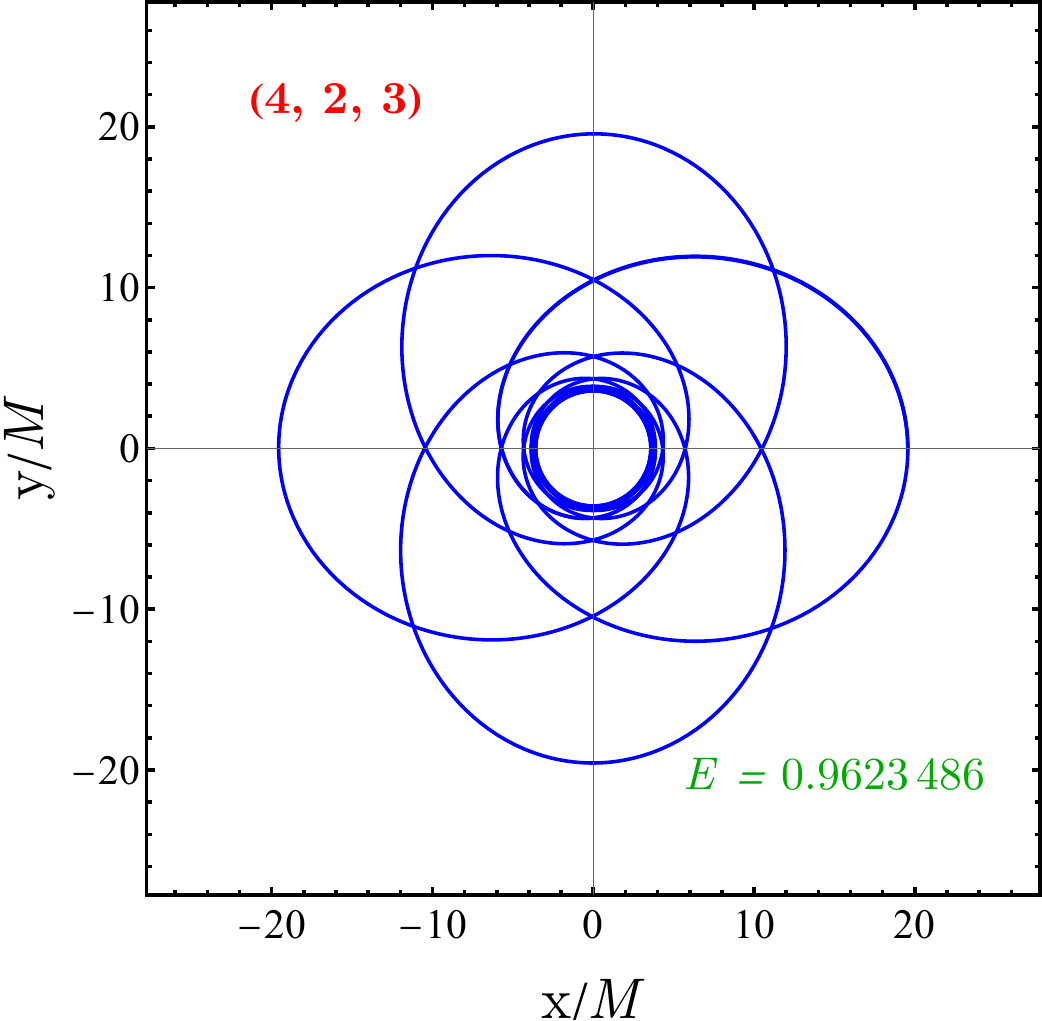}
    \end{tabular}
    
    \caption{Representative periodic orbits in the \textit{NC Schwarzschild}
spacetime, classified by the triplet $(z,w,v)$. The angular momentum is fixed at
$L_{\mathrm{av}}$, while the energy required for each orbit is indicated in the
corresponding panel. The \textit{NC} parameter is set to
$\Theta/M^2=0.02$.}
    
    \label{fig:allorbits}
\end{figure}

The corresponding orbital trajectories are displayed in Fig.~\ref{fig:allorbits} for representative periodic orbits around the \textit{NC Schwarzschild} black hole. Each orbit is characterized by a triplet $(z,w,v)$, where $z$ determines the number of leaves, $w$ gives the number of whirls around the central object, and $v$ specifies the order in which successive leaves are traced. As expected, increasing the zoom number $z$ produces more intricate orbital patterns. By contrast, larger values of the whirl number $w$ increase the number of revolutions performed near the black hole between two consecutive apoapsis passages.

\begin{table}[htbp]
\centering
\small
\setlength{\tabcolsep}{3pt}
\caption{Energy values $E$ for selected periodic orbits characterized by
$(z,w,v)$ in the \textit{NC Schwarzschild} spacetime. The angular momentum
is fixed at $L_{\mathrm{av}}$, while the \textit{NC} parameter
$\Theta/M^2$ is varied.}
\label{tab:qE}

\begin{tabular}{|c|S[table-format=1.5]|*{8}{S[table-format=1.6]|}}
\hline
{$\Theta/M^2$} & {$L_{\mathrm{av}}$} &
{$E(1,1,0)$} &
{$E(2,1,1)$} &
{$E(3,1,2)$} &
{$E(4,1,3)$} &
{$E(1,2,0)$} &
{$E(2,2,1)$} &
{$E(3,2,2)$} &
{$E(4,2,3)$} \\
\hline \hline
0.00 & 3.73205 & 0.965425 & 0.968026 & 0.968225 & 0.968285 & 0.968383 & 0.968434 & 0.968438 & 0.968440 \\ \hline
0.01 & 3.48813 & 0.960789 & 0.964022 & 0.964281 & 0.964362 & 0.964493 & 0.964566 & 0.964572 & 0.964574 \\ \hline
0.02 & 3.36904 & 0.957966 & 0.961679 & 0.961988 & 0.962084 & 0.962238 & 0.962338 & 0.962344 & 0.962348 \\ \hline
0.03 & 3.26636 & 0.955102 & 0.959384 & 0.959756 & 0.959874 & 0.960073 & 0.960191 & 0.960201 & 0.960204 \\
\hline
\end{tabular}
\end{table}


\section{Precession of Nearly Periodic Orbits}

The orbital parameter $q$ provides a useful way to classify bound trajectories. Closed periodic orbits are associated with rational values of $q$, whereas generic bound motion generally corresponds to irrational values and therefore does not repeat itself after a finite number of radial cycles. Since nearby rational ones can approximate irrational numbers, a generic orbit may be viewed as a small departure from an appropriate periodic orbit~\cite{healy2009zoom}. This idea is particularly useful for describing precessing trajectories close to exact closed solutions.

\begin{figure}[ht!]
\centering
\includegraphics[width=55mm]{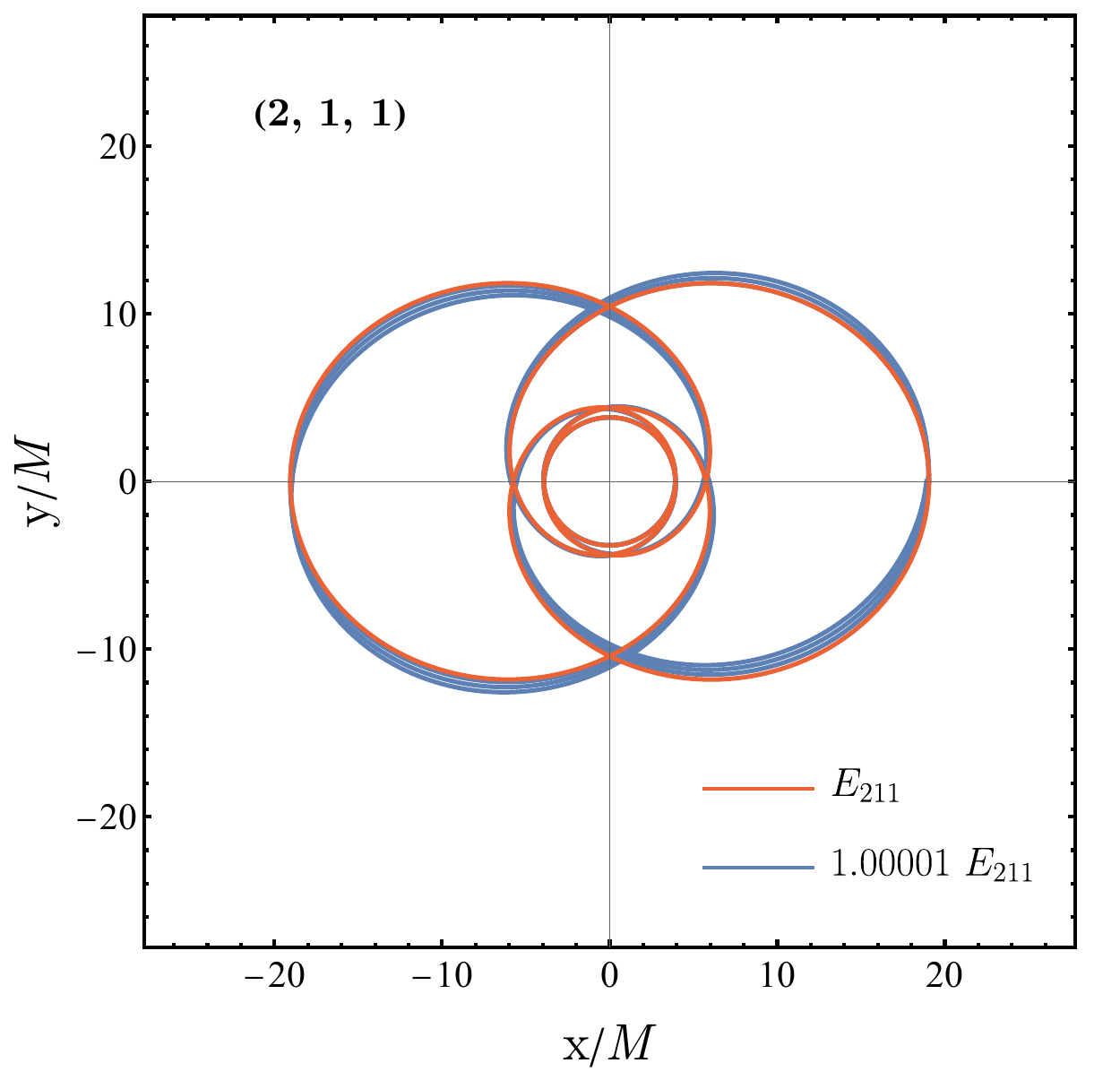}
\includegraphics[width=55mm]{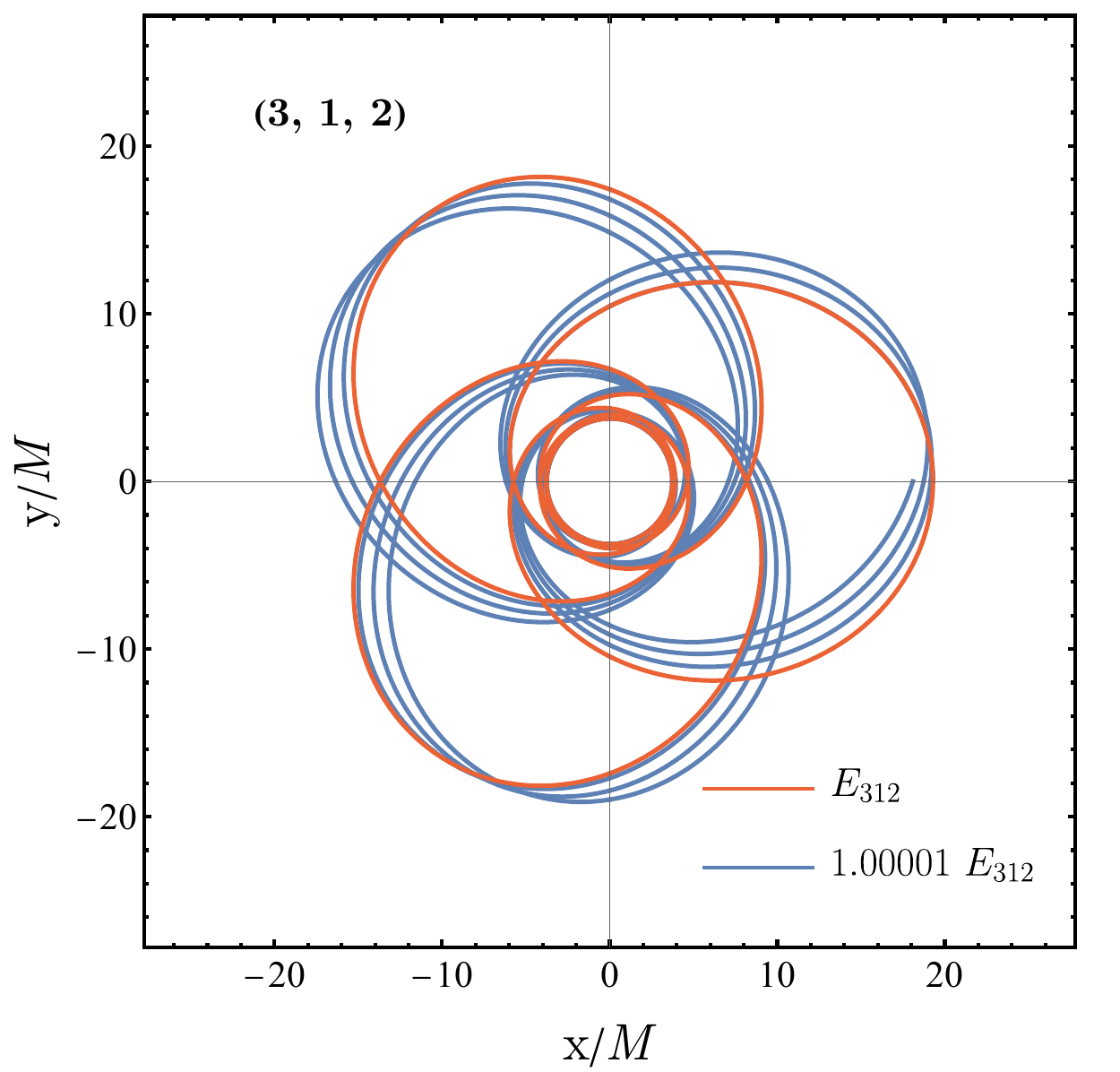}
\includegraphics[width=55mm]{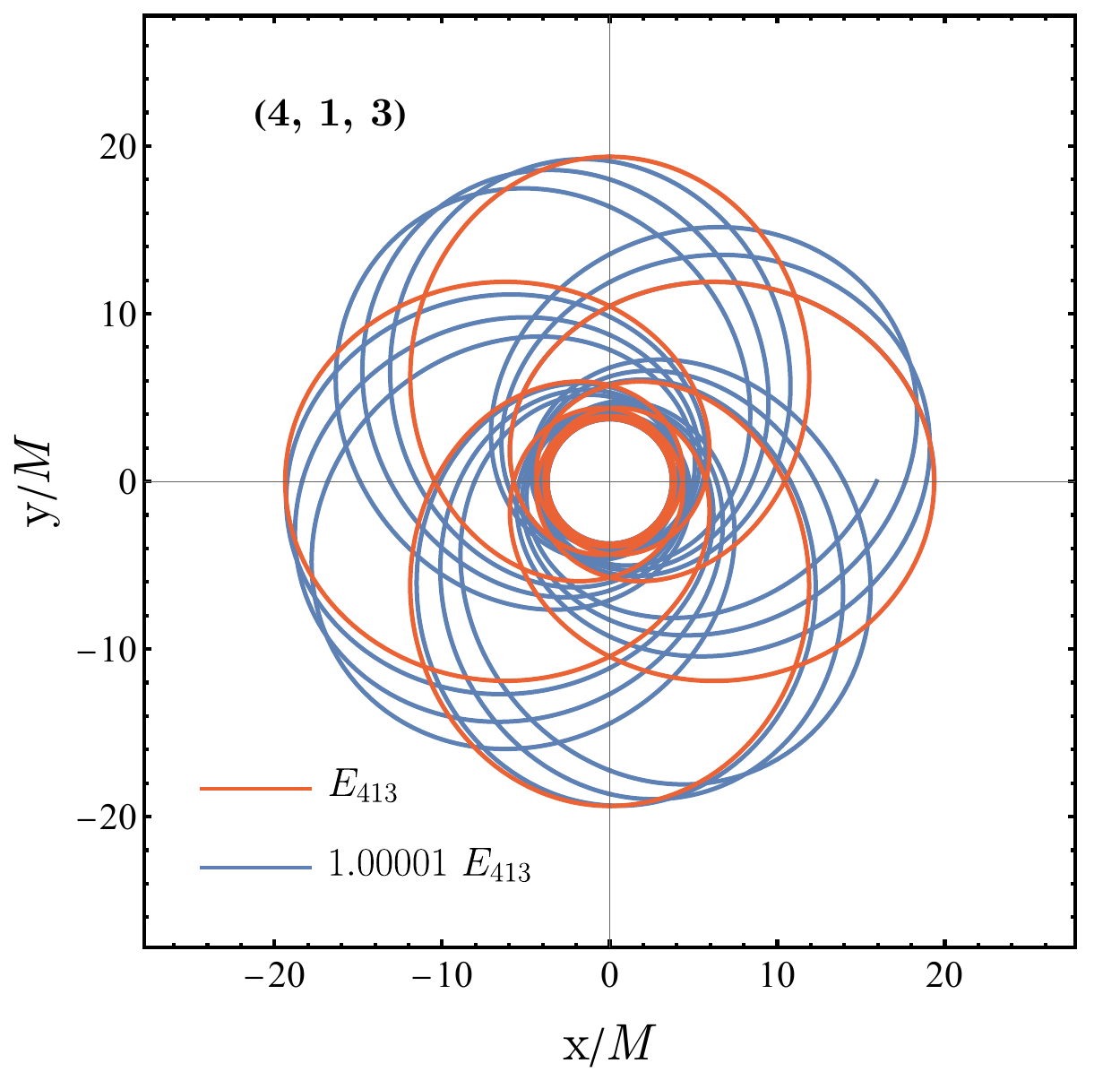}
\caption{Exact periodic orbits and nearby precessing trajectories for different $(z,\omega,v)$ with $\Theta/M^2=0.02$. The red curves show the closed periodic solutions, whereas the blue curves correspond to the trajectories of the relative energy shift $\Delta E/E_0=0.0001$, with the angular momentum held fixed at $L_{\mathrm{av}}$.}
\label{fig:prec1}
\end{figure}

Following Refs.~\cite{wang2022periodic,shabbir2025periodic}, we write the orbital parameter of a nearly periodic trajectory as
\begin{equation}
q = \omega+\frac{v}{z}\pm\delta ,
\end{equation}
where $z$, $\omega$, and $v$ specify the reference periodic orbit, and $0<\delta\ll1$ denotes the departure from the exact rational value. Once this small deviation is present, the trajectory no longer closes on itself.
We consider three representative periodic configurations,
$(2,1,1)$, $ (3,1,2)$, and $(4,1,3)$ with $\Theta/M^2=0.02$. For each case, the angular momentum is fixed at the corresponding average value $L_{\mathrm{av}}$, while the energy is shifted slightly from its periodic value, with $\Delta E/E_0=0.00001$. The resulting values of $q$ are then used to determine the corresponding deviations from exact periodicity. The results are listed in Table~\ref{tab:delta}. The results show that even a subtle change in the energy of the particle leads to irrational number with $\delta$ deviation. The trace of this deviation on the orbits are demonstrated in Fig.~\ref{fig:prec1}, which shows the exact periodic trajectories together with their nearby precessing counterparts. 

\begin{table}[!ht]
\centering
\caption{Periodic energy $E_0$, relative energy shift $\Delta E/E_0$, and deviation parameter $\delta$ for the nearly periodic trajectories shown in Fig.~\ref{fig:prec1}. The angular momentum is fixed at $L_{\mathrm{av}}$ in all cases.}
\setlength{\tabcolsep}{10pt}
\begin{tabular}{|c|c|c|c|c|}
\hline
$(z,\omega,v)$ & $E_0$ & $\Delta E/E_0$ & $q_0$ & $\delta$ \\
\hline \hline
$(2,1,1)$ & $0.961679$ & $0.00001$ & $3/2$ & $0.0039$ \\
\hline
$(3,1,2)$ & $0.961999$ & $0.00001$ & $5/3$ & $0.0072$ \\
\hline
$(4,1,3)$ & $0.962084$ & $0.00001$ & $7/4$ & $0.0099$ \\
\hline
\end{tabular}\label{tab:delta}
\end{table}

The red curves correspond to the closed periodic solutions, while the blue curves represent the trajectories obtained after the small energy shift. To ensure a uniform comparison among the three cases, the total angular interval is fixed to $50\pi$. 

The comparison shows that even a small energy variation can move the orbit away from the rational periodic condition and convert a closed trajectory into a slowly precessing one. The values of $\delta$ reported in Table~\ref{tab:delta} further indicate that this response is not the same for all configurations, although the same relative energy shift is applied in each case. This behaviour is also visible in Fig.~\ref{fig:prec1}, where the accumulated deviation is largest for $(4,1,3)$, followed by $(3,1,2)$ and then $(2,1,1)$. This suggests that the degree of precessional drift is so sensitive to the specific periodic configuration.


\section{Preliminary bound of the precession}

The relativistic precession of stellar orbits around Sgr~A$^*$ provides a useful probe of possible deviations from the \textit{Schwarzschild} geometry in the weak--field regime \cite{abuter2020detection,yao2026probing, li2024precessing,lin2022precessing}. In particular, the high--precision astrometric and spectroscopic observations of the S2 star by the GRAVITY Collaboration have confirmed the \textit{Schwarzschild} precession predicted by general relativity within the current observational uncertainty. This result can therefore be used to place a preliminary bound on the \textit{NC} parameter of the spacetime considered in this work.
It is convenient to introduce the parameter
\begin{equation}
\beta \equiv \frac{8\sqrt{\Theta}}{\sqrt{\pi}p},
\end{equation}
where
\begin{equation}
p\equiv a(1-e^2)
\end{equation}
is the semi--latus rectum of the orbit, and $a$ and $e$ denote the semi--major axis and eccentricity, respectively. With this notation, the metric function in Eq.~\eqref{fNC} may be written as
\begin{equation}
f(r)=1-\frac{2M}{r}+\frac{Mp\beta}{r^2}.
\end{equation}
The Keplerian parametrization of the radial coordinate is taken as
\begin{equation}
r(\psi)=\frac{p}{1+e\cos\psi}
=\frac{a(1-e^2)}{1+e\cos\psi},
\end{equation}
 while $\psi$ is the relativistic anomaly. The periastron and apastron are then given by
\begin{equation}
r_p=\frac{p}{1+e}=a(1-e), \qquad
r_a=\frac{p}{1-e}=a(1+e).
\end{equation}

\begin{equation}
\frac{\mathrm{d}\phi}{\mathrm{d}\psi}
=\frac{Le\sin\psi}{p\sqrt{ E^2-f(r(\psi))\left[1+\frac{L^2}{r^2(\psi)}\right] }}.
\end{equation}
The constants $E$ and $L$ are fixed by imposing $\dot r=0$ in Eq.~\ref{eq:rdot} at the two radial turning points which gives
\begin{equation}
E^2= \frac{ f(r_a)f(r_p)(r_p^2-r_a^2) }{ r_p^2 f(r_a)-r_a^2 f(r_p) },
\end{equation}
and
\begin{equation}
L^2= \frac{ r_a^2r_p^2\left[f(r_p)-f(r_a)\right] }{ r_p^2 f(r_a)-r_a^2 f(r_p) }.
\end{equation}

Substituting these expressions into $d\phi/d\psi$ and expanding in the weak--field regime
\begin{equation}
\frac{M}{p}\ll 1, \qquad \beta\ll 1,
\end{equation}
one obtains, to leading order,
\begin{equation}
\frac{\mathrm{d}\phi}{\mathrm{d}\psi}
\simeq 1+ \frac{(3+e\cos\psi)M}{p} -\frac{\beta}{2}.
\end{equation}
The first term corresponds to the Newtonian closed orbit, the second term gives the standard \textit{Schwarzschild} contribution, and the last term represents the leading \textit{NC} correction.

The total azimuthal angle accumulated during one radial period is therefore
\begin{equation}
\Delta\phi \simeq 2\int_0^\pi \left[ 1+ \frac{(3+e\cos\psi)M}{p} -\frac{\beta}{2} \right]\mathrm{d}\psi.
\end{equation}
Carrying out the integration gives
\begin{equation}
\Delta\phi \simeq 2\pi+\frac{6\pi M}{p}-\pi\beta .
\end{equation}
The periastron advance is obtained by subtracting the Newtonian contribution $2\pi$,
\begin{equation}
\Delta\omega_{\rm NC} \equiv \Delta\phi-2\pi \simeq \frac{6\pi M}{p}-\pi\beta .
\end{equation}
Substituting the definition of $p$ and $\beta$, this can be written as
\begin{equation}
\Delta\omega_{\rm NC} \simeq \frac{6\pi M}{a(1-e^2)} -\frac{8\sqrt{\pi\Theta}}{a(1-e^2)} .
\end{equation}
Equivalently,
\begin{equation}
\Delta\omega_{\rm NC} \simeq \Delta\omega_{\rm GR} + \Delta\omega_{\Theta},
\end{equation}
where
\begin{equation}
\Delta\omega_{\rm GR} =\frac{6\pi M}{a(1-e^2)}, \qquad \Delta\omega_{\Theta} =-\frac{8\sqrt{\pi\Theta}}{a(1-e^2)} .
\end{equation}

\begin{figure}[ht!]
\centering
\includegraphics[width=85mm]{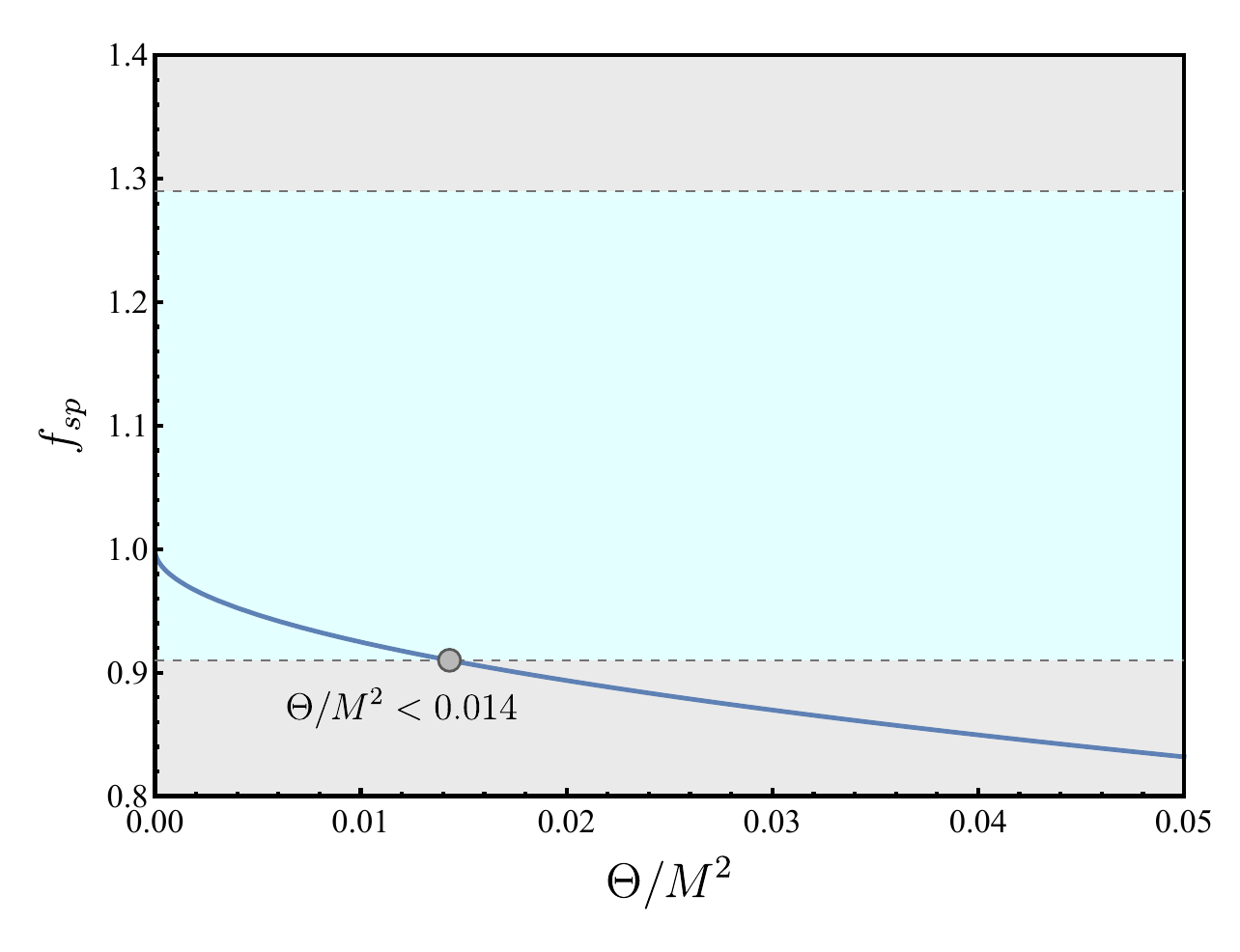}
\caption{Precession ratio $f_{sp}^{\rm NC}$ as a function of the dimensionless \textit{NC} parameter $\Theta/M^2$. The blue region shows the observationally allowed range from the S2 star precession measurement.}
\label{fig:prec}
\end{figure}

To compare this result with the observed \textit{Schwarzschild} precession of the S2 star, we define
\begin{equation}
f_{sp}^{\rm NC} \equiv \frac{\Delta\omega_{\rm NC}}{\Delta\omega_{\rm GR}},
\end{equation}
which leads to
\begin{equation}
f_{sp}^{\rm NC} \simeq 1- \frac{4\sqrt{\Theta}}{3\sqrt{\pi}M}.
\end{equation}
This expression shows that the \textit{NC} correction decreases the predicted precession relative to the \textit{Schwarzschild} value. Using the observational parameters of the S2 orbit around Sgr~A$^*$, we evaluate the ratio $f_{sp}^{\rm NC}$
as a function of the dimensionless \textit{NC} parameter $\Theta/M^2$. The observationally allowed interval inferred from the GRAVITY measurement is \cite{abuter2020detection}
\begin{equation}
1.10-0.19 \leq f_{sp} \leq 1.10+0.19 .
\end{equation}
which is shown by the green band in Fig.~\ref{fig:prec}, and the solid curve represents the prediction of the \textit{NC Schwarzschild} spacetime.
As the \textit{NC} parameter increases, the predicted precession ratio decreases below the \textit{Schwarzschild} value. Requiring the theoretical curve to remain inside the observational band leads to the preliminary bound
\begin{equation}
\frac{\Theta}{M^2}<0.014 .
\end{equation}
Thus, the current S2 star precession data put an upper bound on the allowed strength of the \textit{NC} correction.

\section{Gravitational waveform from periodic orbits}

Extreme Mass Ratio Inspirals (EMRIs), consisting of a stellar mass compact object orbiting a supermassive black hole, are among the most important targets of space--based gravitational wave detectors such as LISA, Taiji, and TianQin \cite{amaro2017laser,hu2017taiji,gong2021concepts}. The gravitational waves emitted by these systems encode detailed information about the orbital dynamics and the geometry of the central object.
The \textit{NC} corrections modify the spacetime geometry and consequently affect the orbital motion of the particle. These changes can leave observable imprints on the emitted gravitational wave signal, allowing the waveform to serve as a potential probe of the non--commutative structure of spacetime.
Our analysis is performed within the approximation \cite{hughes2000evolution,hughes2001evolution,hughes2005gravitational,sundararajan2007towards,miller2021two}, which is appropriate for EMRIs since the radiation reaction timescale is much longer than the orbital period. In this regime, the orbital energy and angular momentum vary slowly and can be treated as constants over several orbital cycles. Since we focus on short term orbital evolution, the back reaction of gravitational wave emission on the orbit is neglected \cite{isoyama2022adiabatic}.

To compute the waveforms, we adopt the numerical kludge approach \cite{babak2007kludge}, a widely used framework for modeling EMRI signals . The method proceeds in two steps. First, the geodesic equations are numerically integrated to obtain the periodic trajectories presented in the previous section. These trajectories are then used as input for the quadrupole formula of gravitational radiation to generate the corresponding waveforms \cite{yang2025gravitational,heidari2026gravitational}. 
The radiation field is described by the symmetric trace--free (STF) mass quadrupole tensor, and the corresponding metric perturbation in the wave zone is given by~\cite{zhao2025periodic}
\begin{eqnarray}
h_{ij} = \frac{4\mu M}{D_L} \left( v_i v_j - \frac{m}{M}n_i n_j \right),
\label{eq:quadrupole}
\end{eqnarray}
where $M$ and $m$ denote the masses of the central black hole and the orbiting compact object, respectively, $D_L$ is the luminosity distance to the source, and $\mu=\frac{Mm}{(M+m)^2}$, is the symmetric mass ratio. The quantities $n_i$ and $v_i$ represent the radial unit vector and the velocity components of the orbiting body. Eq.~(\ref{eq:quadrupole}) captures the dominant quadrupolar contribution to the gravitational radiation emitted by the system. To obtain the observable waveform, the metric perturbation is projected onto a transverse--traceless frame adapted to the detector. Following Refs.~\cite{poisson2014gravity,yang2025gravitational,zhao2025periodic}, we introduce the orthonormal basis vectors
\begin{eqnarray}
e_X &=& (\cos\zeta,-\sin\zeta,0), \nonumber\\ e_Y &=& (\cos\iota\sin\zeta,\cos\iota\cos\zeta,-\sin\iota), \nonumber\\ e_Z &=& (\sin\iota\sin\zeta,\sin\iota\cos\zeta,\cos\iota),
\label{eq:basis}
\end{eqnarray}
where $\iota$ denotes the inclination angle and $\zeta$ specifies the orientation of the orbit relative to the observer. The two independent polarization modes are then expressed as~\cite{maselli2022detecting,liang2023probing}
\begin{eqnarray}
h_{+} &=& -\frac{2\mu M^2}{D_L r} \left(1+\cos^2\iota\right) \cos(2\phi+2\zeta), \label{eq:hplus} \\ h_{\times} &=& -\frac{4\mu M^2}{D_L r} \cos\iota \sin(2\phi+2\zeta),
\label{eq:hcross}
\end{eqnarray}
with $\phi$ denoting the orbital phase. These expressions describe the plus and cross polarizations of the gravitational wave signal measured by a distant observer.

Throughout this analysis, geometrized units ($G=c=1$) are adopted. Unless otherwise specified, the EMRI system consists of a $10\,M_\odot$ compact object orbiting a $10^7\,M_\odot$ \textit{NC Schwarzschild} black hole. The source is placed at a luminosity distance of $D_L=200\,\mathrm{Mpc}$, corresponding to $\approx 4.18\times10^{14}$ in units of the black hole mass. The orientation angles are fixed at $\iota=\zeta=\pi/4$. The waveforms are generated over one complete cycle of the periodic orbit, with the phase evolution obtained from the numerical integration of the geodesic equations.

\begin{figure*}[ht!]
	\centering
	\begin{minipage}{0.48\textwidth}
		\centering
		\includegraphics[width=\textwidth]{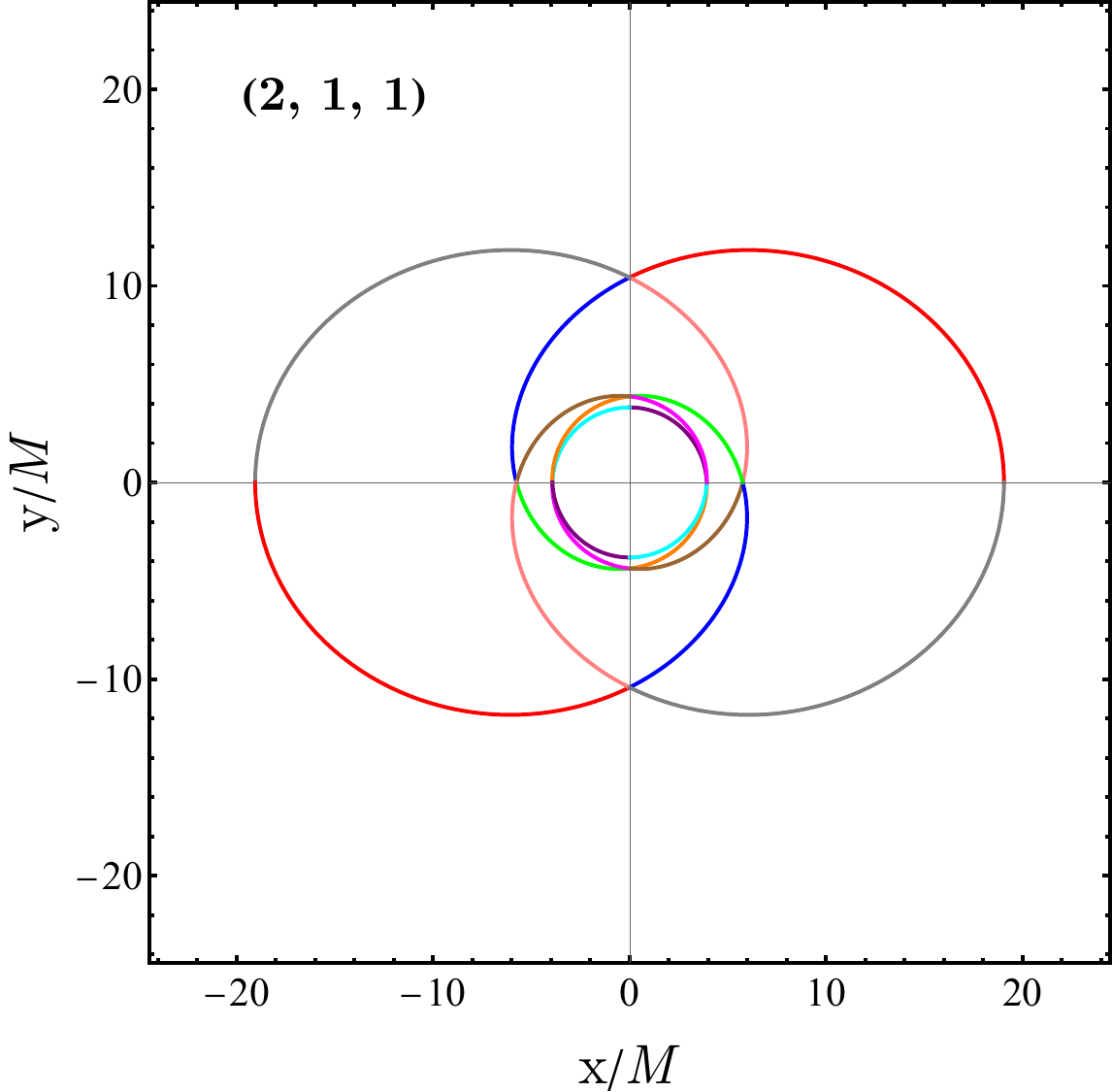}
	\end{minipage}
	\hfill
	\begin{minipage}{0.48\textwidth}
		\centering
		\includegraphics[width=\textwidth]{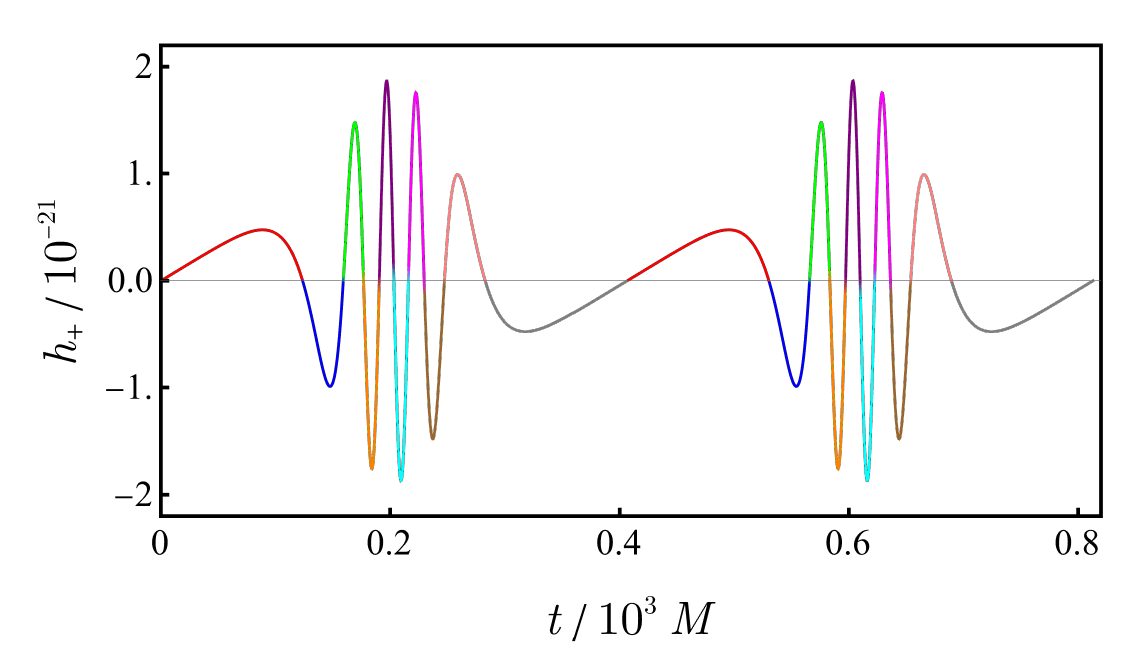}
		\includegraphics[width=\textwidth]{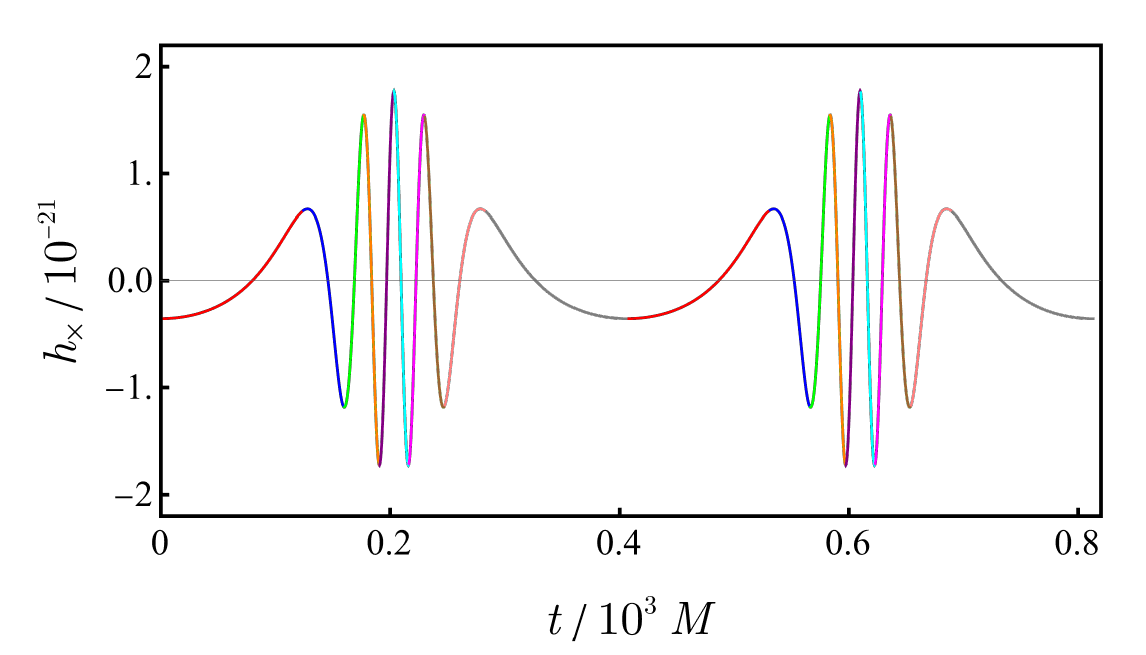}
	\end{minipage}
	\caption{The $(2,1,1)$ periodic orbit in an EMRI system around a \textit{NC Schwarzschild} black hole characterized by the parameter $\Theta/M^2=0.02$, together with the corresponding gravitational wave signals. Different portions of the trajectory are distinguished by separate color schemes.}
	\label{fig:colored}
\end{figure*}

 Fig.~\ref{fig:colored} displays the $(2,1,1)$ periodic orbit together with the corresponding gravitational wave polarizations $h_{+}$ and $h_{\times}$. To facilitate the interpretation of the waveform, the orbit is partitioned into consecutive phase intervals of width $\pi/2$, with the same color coding applied to the associated segments of the gravitational wave signal. This representation establishes a direct correspondence between specific regions of the trajectory and their imprint on the emitted radiation.
A pronounced variation in the waveform amplitude is observed over a single orbital cycle. The largest contributions arise during the phases in which the particle traverses the innermost portion of the orbit, where the gravitational field is strongest and the orbital velocity reaches its maximum. In contrast, the signal amplitude decreases as the particle moves toward larger radii, reflecting the weaker dynamical influence of the central object. As a result, both polarization modes exhibit a sequence of enhanced bursts interspersed with lower amplitude intervals, mirroring the non-uniform distribution of the orbital motion.
The phase--dependent structure of the waveform highlights the close connection between the orbital dynamics and the emitted gravitational radiation. Since the waveform is determined by the underlying particle motion, any modification of the spacetime geometry is expected to leave a corresponding imprint on the radiative signal. Periodic EMRI waveforms therefore provide a useful probe of \textit{NC} corrections and their influence on the strong--field dynamics.

\begin{figure}[ht!]
	\centering
	\begin{minipage}{0.48\textwidth}
		\centering
		\includegraphics[width=\textwidth]{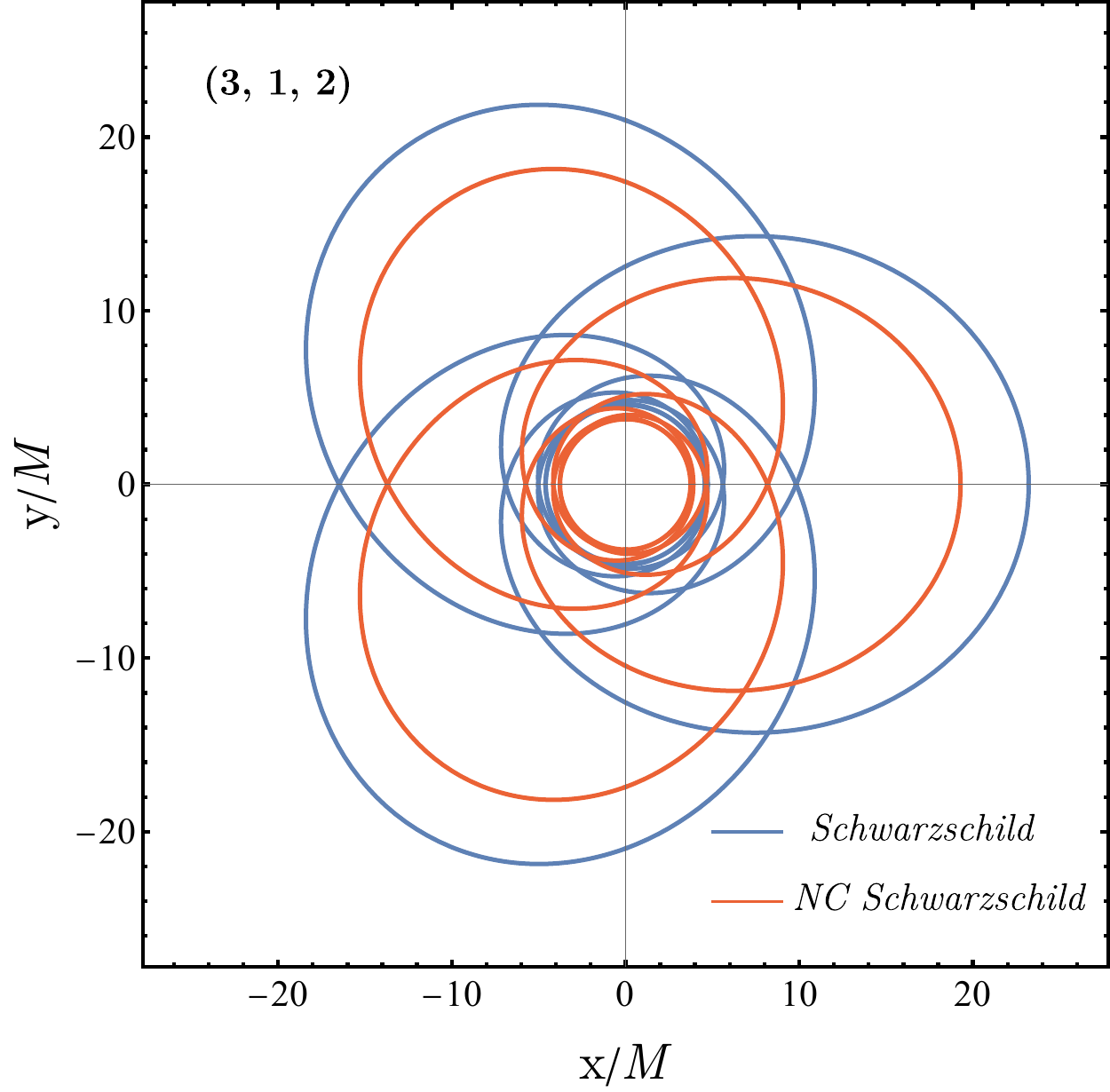}
	\end{minipage}
	\hfill
	\begin{minipage}{0.48\textwidth}
		\centering
		\includegraphics[width=\textwidth]{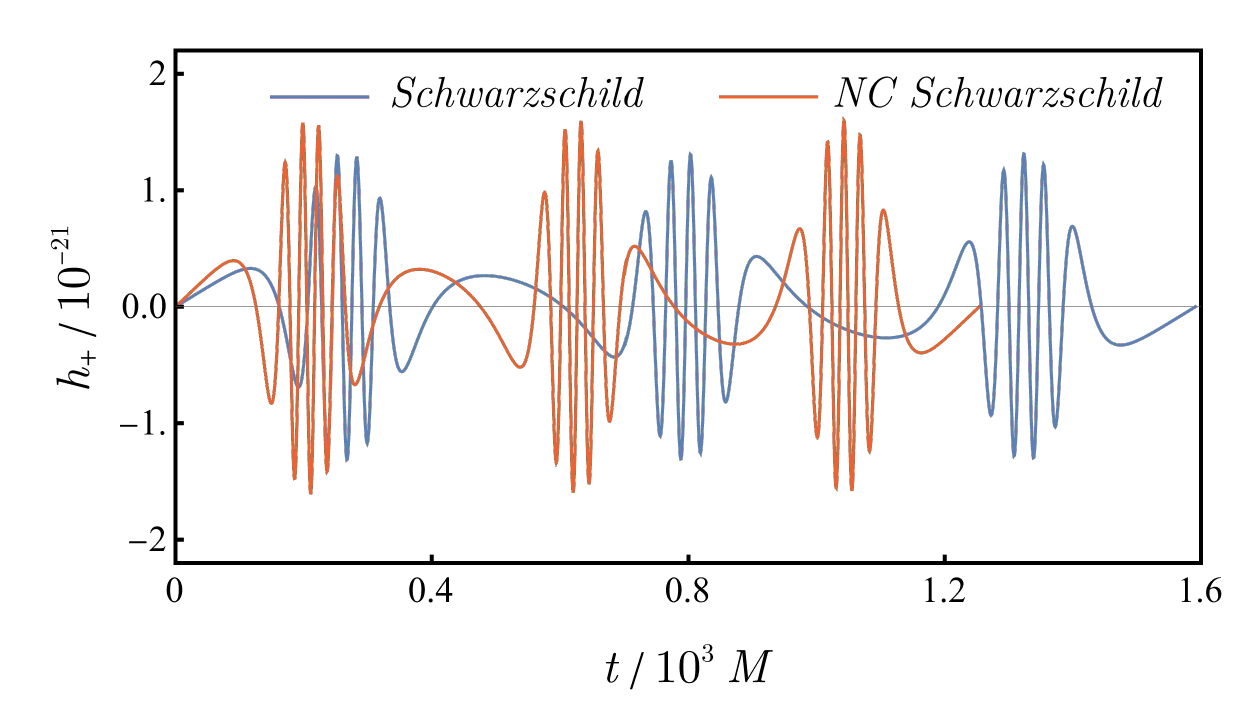}\\[0.5cm]
		\includegraphics[width=\textwidth]{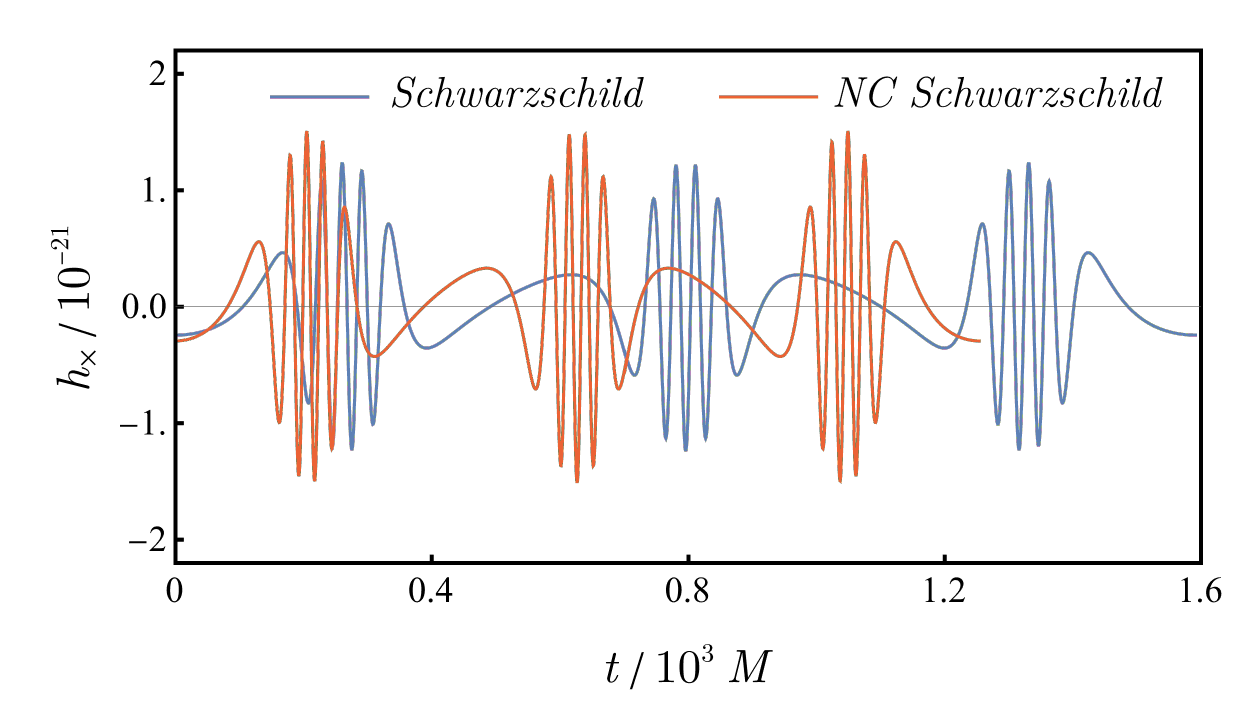}
	\end{minipage}
	\caption{Orbital trajectories (left) and the corresponding gravitational wave polarizations $h_+$ (top right) and $h_\times$ (bottom right) for a \textit{Schwarzschild} black hole in commutative (blue) and non--commutative (red) spacetimes. For the latter, the \textit{NC} parameter is fixed at $\Theta/M^2=0.02$.}
	\label{fig:comp}
\end{figure}

Fig.~\ref{fig:comp} presents a representative comparison between periodic motion in the commutative \textit{Schwarzschild} spacetime and its non--commutative counterpart. While the overall topology of the orbit remains unchanged, the introduction of \textit{NC} corrections leads to visible shifts in the trajectory, particularly in the vicinity of the central black hole where the gravitational field is strongest. These orbital modifications are directly reflected in the corresponding gravitational wave polarizations, producing differences in both the amplitude and phase evolution of the $h_{+}$ and $h_{\times}$ modes. The comparison demonstrates that the non--commutative deformation affects not only the geometry of the orbit but also the observable properties of the emitted radiation.

\begin{figure}[ht!]
    \centering
    \includegraphics[height=57mm]{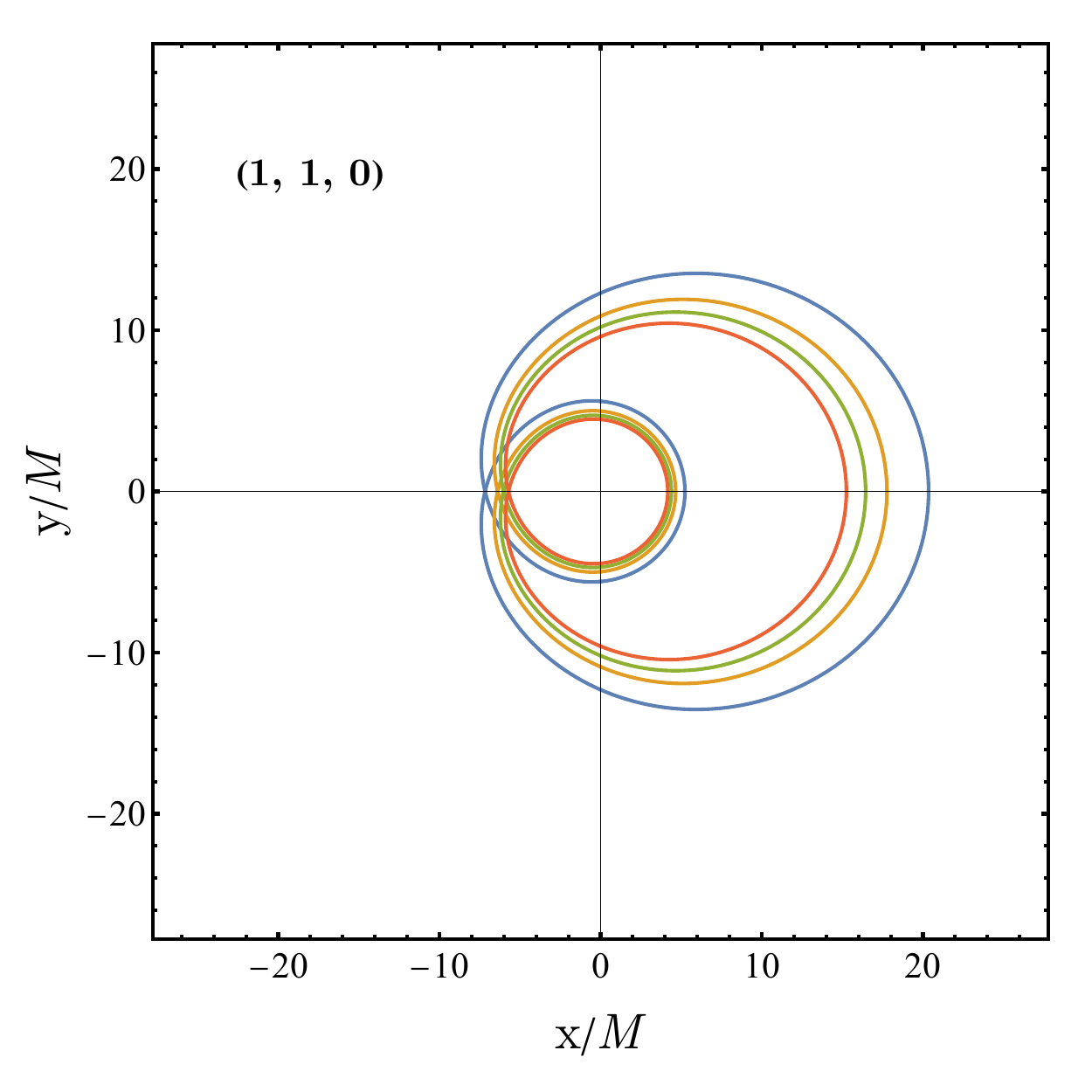}
    \includegraphics[height=57mm]{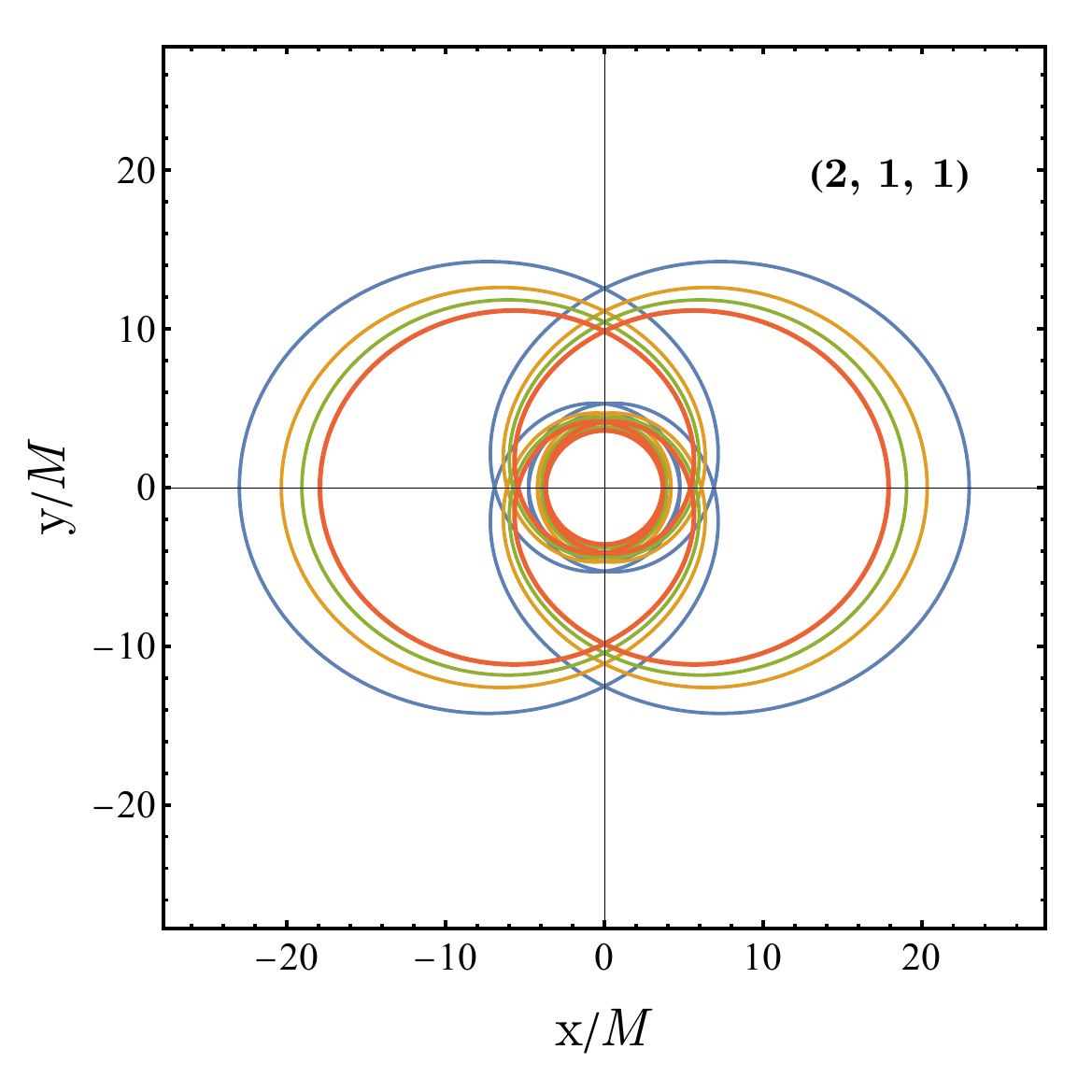}
    \includegraphics[height=57mm]{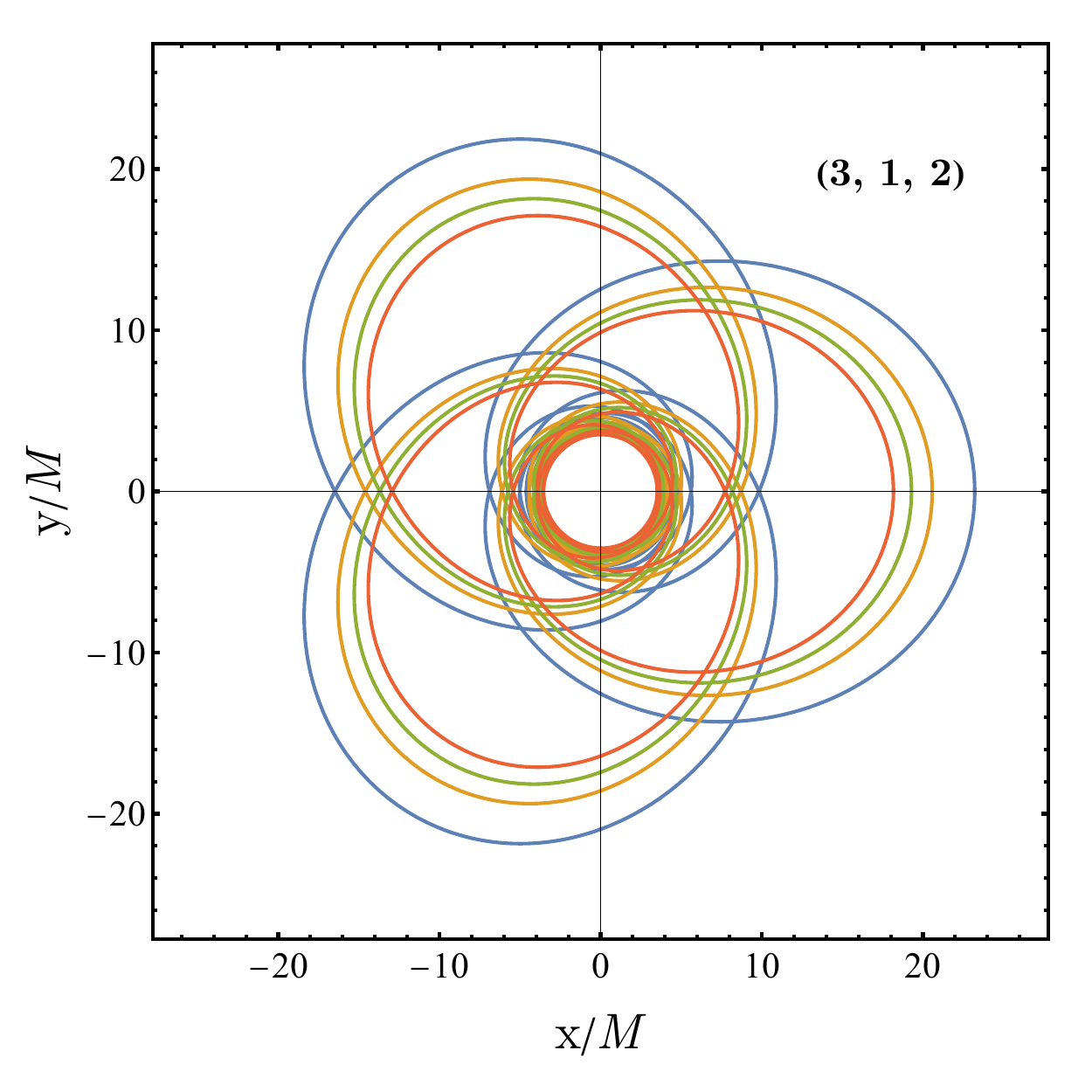}
    \caption{Periodic orbits for three $(z,w,v)$ configurations in \textit{NC Schwarzschild} spacetime. The colors correspond to different values of the \textit{NC} parameter: $\Theta/M^2=0.00$ (blue), $0.01$ (orange), $0.02$ (green), and $0.03$ (red).}
    \label{fig:comporbits}
\end{figure}

\begin{figure}[ht!]
    \centering
    \includegraphics[height=60mm]{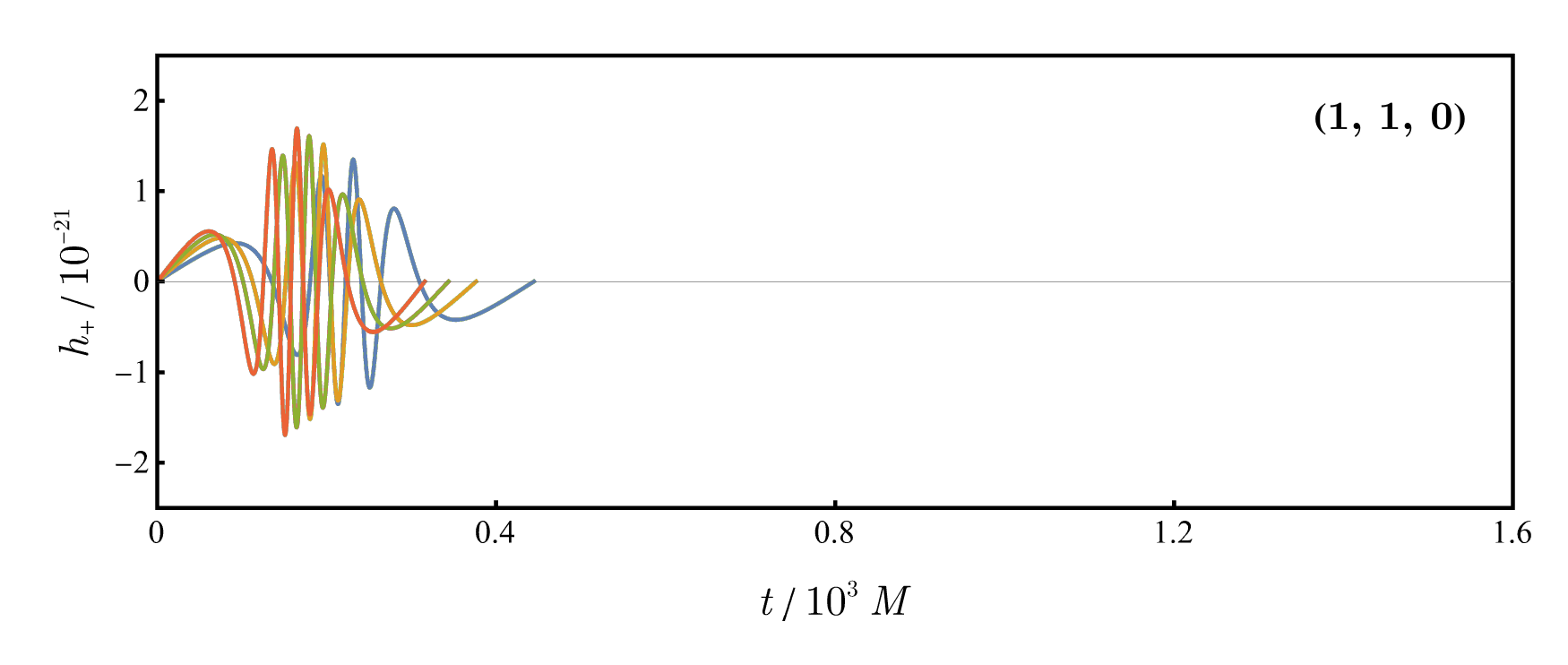}
    \includegraphics[height=60mm]{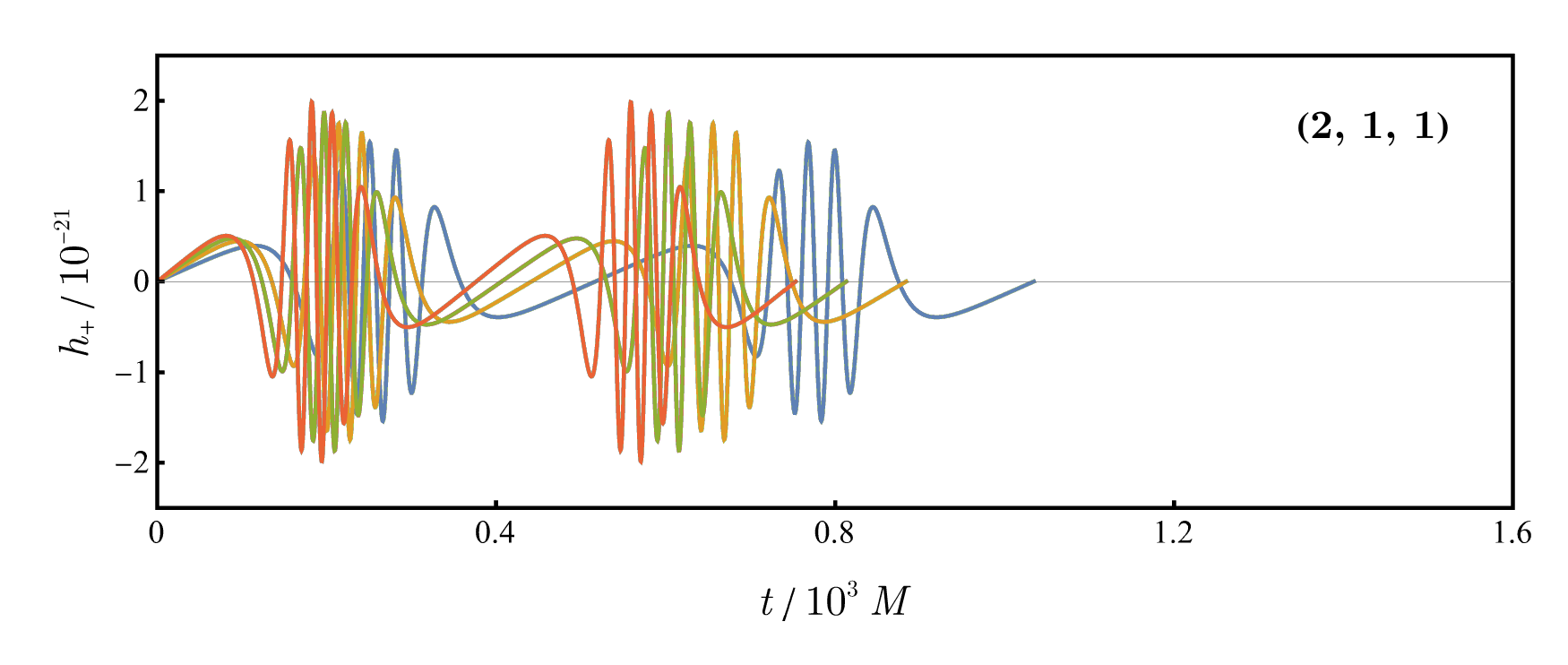}
    \includegraphics[height=60mm]{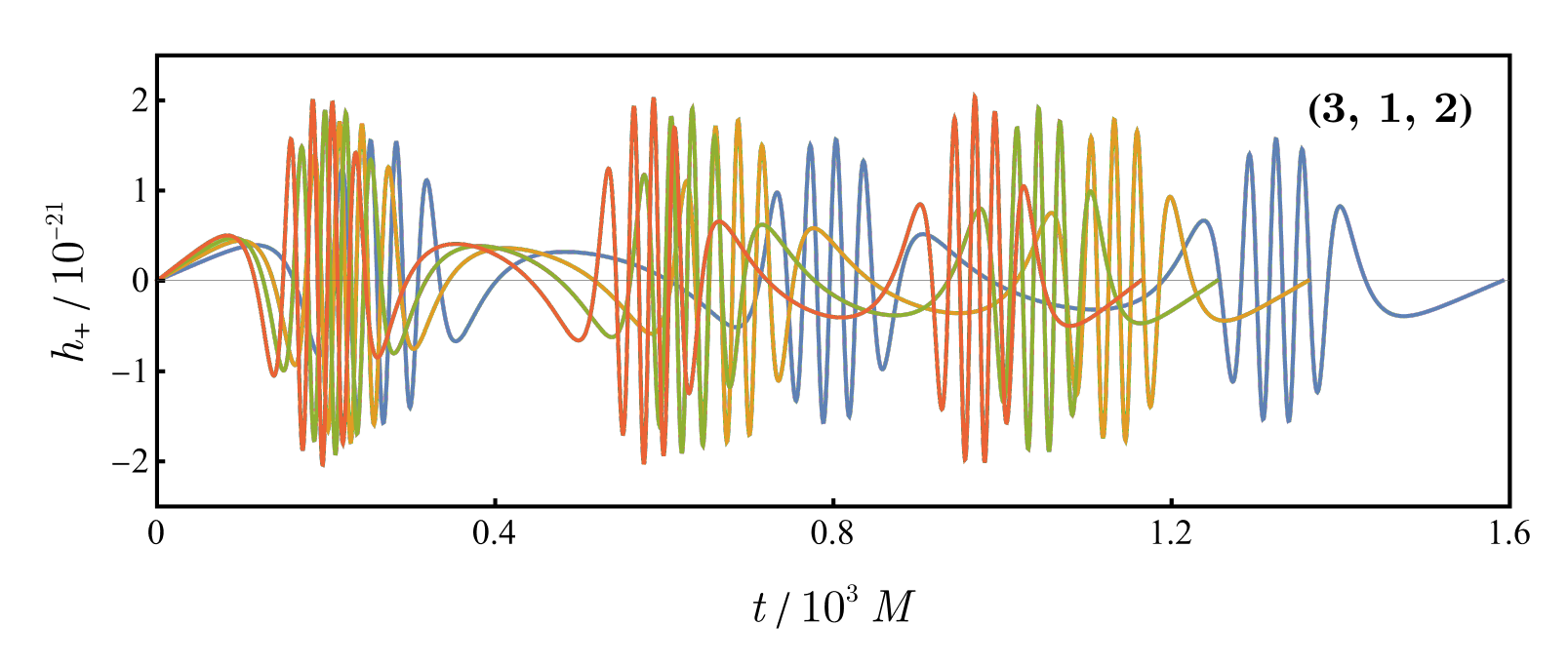}
    \caption{configurations of $(z,w,v)$ in a \textit{NC Schwarzschild} spacetime. The color scheme corresponds to $\Theta/M^2 = 0.00$ (blue), $0.01$ (orange), $0.02$ (green), and $0.03$ (red).} \label{fig:comphpluss}
\end{figure}

\begin{figure}[ht!] 
\centering 
\includegraphics[height=60mm]{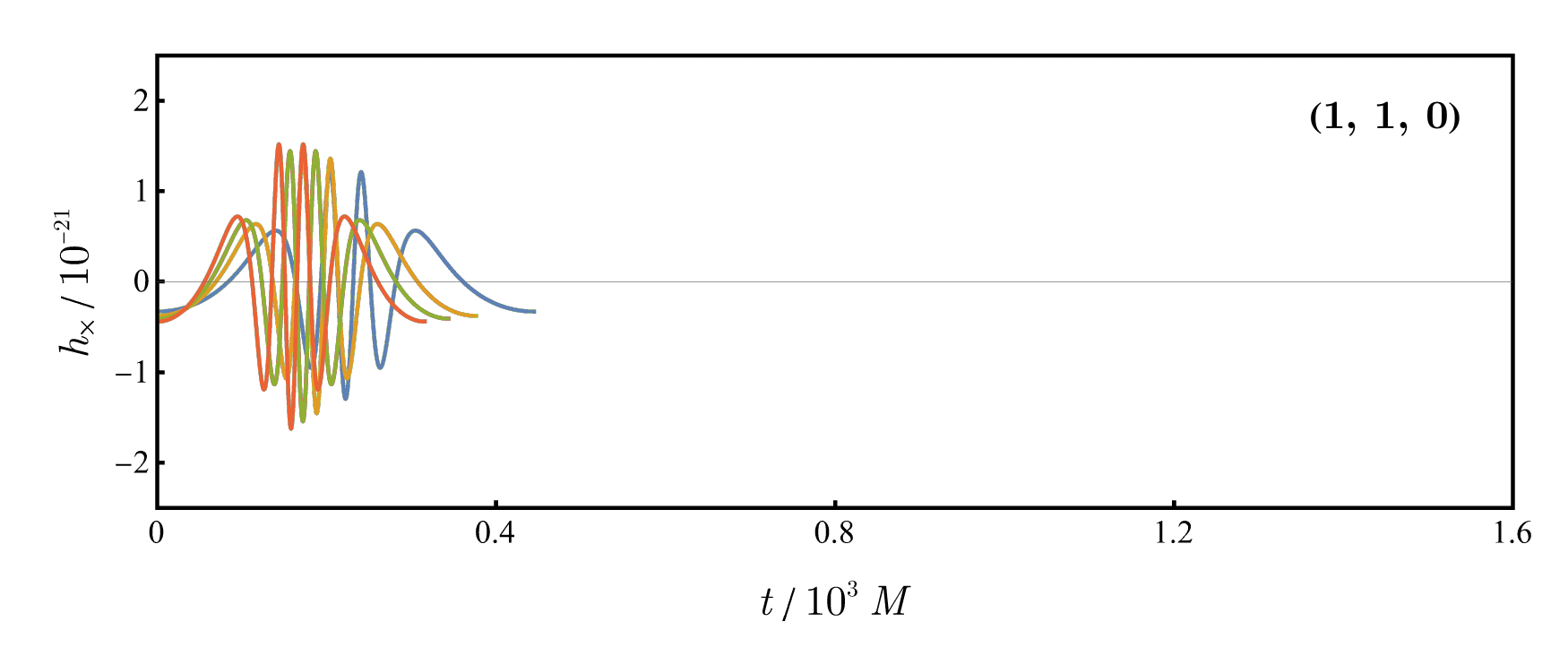} \includegraphics[height=60mm]{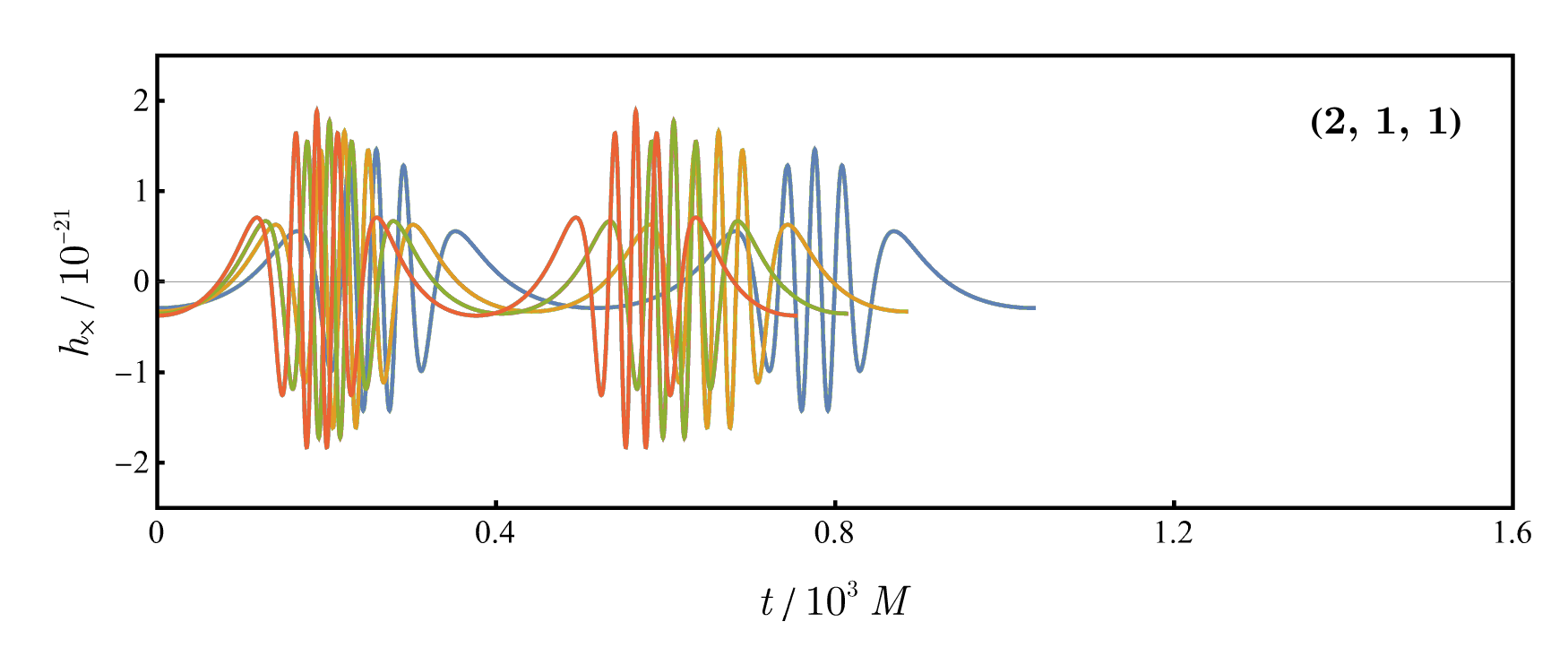} \includegraphics[height=60mm]{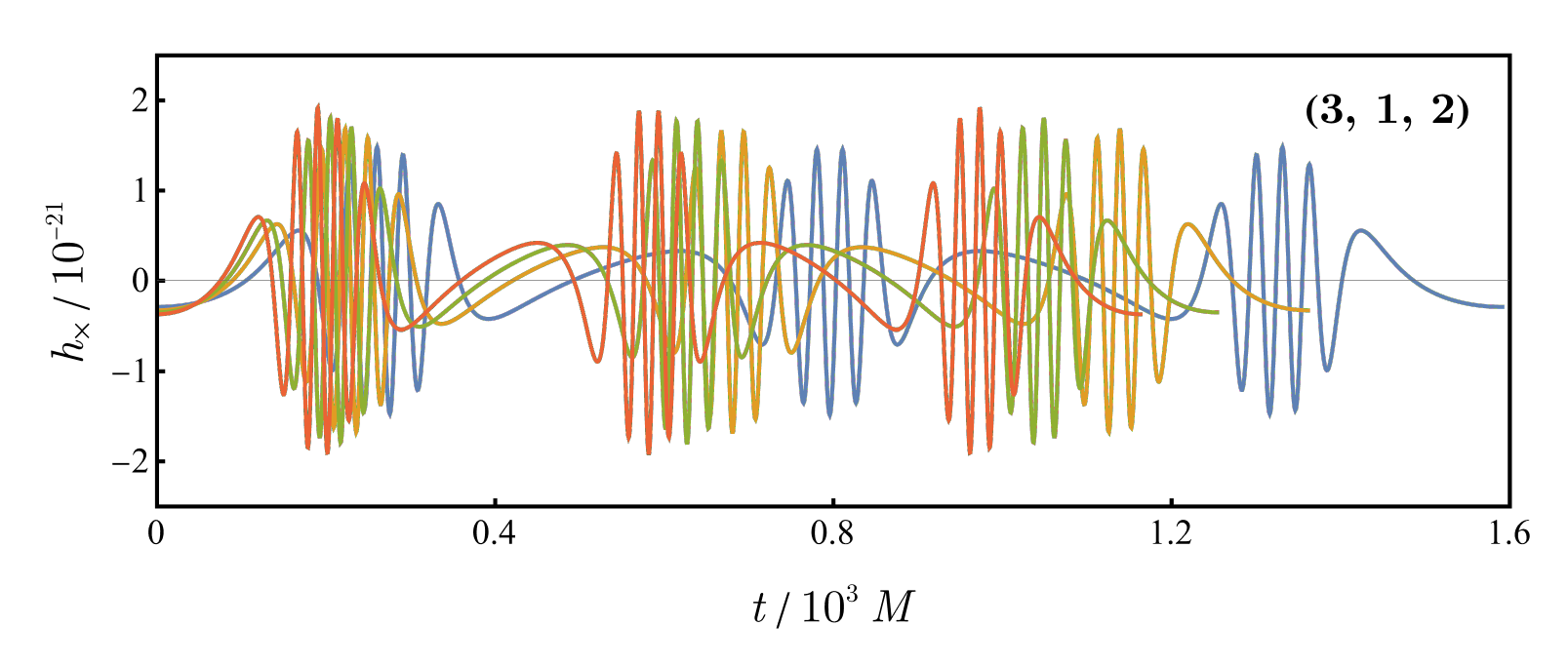} 
\caption{Gravitational wave polarization $h_\times$ for three sets of $(z,w,v)$ configurations in a \textit{NC Schwarzschild} spacetime. The curves are color-coded as $\Theta/M^2 = 0.00$ (blue), $0.01$ (orange), $0.02$ (green), and $0.03$ (red).} \label{fig:comphcrosss} 
\end{figure}

We next investigate how these effects evolve with the scaled deformation parameter $\Theta/M^2$. Figs.~\ref{fig:comporbits}~--~\ref{fig:comphcrosss} display the periodic orbits together with the corresponding gravitational wave polarizations for several values of $\Theta/M^2$ and for three representative configurations $(1,1,0)$, $(2,1,1)$ and $(3,1,2)$.
A common trend can be identified across all configurations. As the \textit{NC} parameter increases, the periodic trajectories become progressively more compact, indicating that the particle explores regions closer to the central black hole. This behavior is accompanied by a systematic enhancement of the gravitational wave signal. In both polarization modes, larger values of $\Theta/M^2$ generally lead to increased amplitudes and noticeable phase shifts relative to the commutative case. Although the quantitative response depends on the orbital configuration, the overall behavior remains qualitatively similar: increasing $\Theta/M^2$ drives the orbit toward smaller characteristic scales and simultaneously amplifies the emitted gravitational radiation. Such signatures suggest that precise measurements of EMRI waveforms could, in principle, provide a great tool to probe the \textit{NC} corrections to the \textit{Schwarzschild} geometry.

\section{Conclusion}

In this work, we investigated the motion of massive test particles and the gravitational wave emission associated with periodic trajectories around a non--commutative \textit{Schwarzschild} black hole sourced by a Lorentzian matter distribution. Initially, we analyzed the effective potential and the conditions governing bound timelike motion. Particular attention was devoted to the marginally bound orbit and the innermost stable circular orbit. We found that increasing the non--commutative parameter displaced both characteristic orbits toward smaller radii and reduced their corresponding angular momenta. This behavior was also reflected in the allowed region of the $(E,L)$ plane, which moved toward lower values of energy and angular momentum. In other words, the non--commutative deformation allowed more tightly bound configurations to occur closer to the central black hole.

Subsequently, we investigated the periodic sector of the timelike geodesics by employing the rational parameter $q$, which relates the azimuthal and radial frequencies. The periodic trajectories were classified through the triplet $(z,w,v)$, allowing different zoom--whirl configurations to be characterized. For a fixed orbital topology, the energy required to produce the corresponding periodic orbit decreased as the non--commutative parameter increased. In parallel, the trajectories became progressively more compact.

We also considered trajectories lying close to the exact periodic configurations. Small variations in the orbital energy were sufficient to move the system away from the rational periodic condition and generate a gradual precessional drift. The accumulated deviation depended on the particular orbit under consideration, indicating that trajectories with different values of $(z,w,v)$ responded differently to the same relative energy perturbation. Moreover, the weak--field expansion of the periastron advance showed that the non--commutativity contribution reduced the \textit{Schwarzschild} prediction. By comparing this result with the observational interval obtained from the S2 star around Sgr~A$^*$, we derived the preliminary constraint
${\Theta}/{M^{2}}<0.014$.

Finally, within the adiabatic and numerical kludge approximations, we calculated the gravitational wave polarizations $h_{+}$ and $h_{\times}$ generated by representative periodic orbits in the extreme mass--ratio regime. The waveform structure followed directly the nonuniform orbital motion. In particular, the signal was enhanced during the stages in which the orbiting body passed through the innermost region of the trajectory and became weaker at larger radial distances. When compared with the \textit{Schwarzschild} case, the non--commutative geometry produced modifications in both the amplitude and phase evolution of the polarizations. Larger values of $\Theta/M^{2}$ generally resulted in more compact trajectories, stronger amplitudes, and noticeable phase displacements, although the quantitative response depended on the orbital configuration.


\section*{Acknowledgments}
\hspace{0.5cm} 
N. H is supported by the Conselho Nacional de Desenvolvimento Científico e Tecnológico (CNPq), grant No. 152891/2025-0. N. H. also acknowledges the networking support provided by COST Action CA22113 -- Fundamental challenges in theoretical physics (Theory and Challenges), CA21106 -- COSMIC WISPers in the Dark Universe: Theory, astrophysics and experiments (CosmicWISPers), CA21136 -- Addressing observational tensions in cosmology with systematics and fundamental physics (CosmoVerse), and CA23130 -- Bridging high and low energies in search of quantum gravity (BridgeQG). A.A.A.F. is supported by Conselho Nacional de Desenvolvimento Cient\'{\i}fico e Tecnol\'{o}gico (CNPq) with the project number 150223/2025-0. I. P. L. was partially supported by the National Council for Scientific and Technological Development - CNPq, grant 312547/2023-4 and to acknowledge networking support by the COST Action BridgeQG (CA23130), the COST Action RQI (CA23115) and the COST Action FuSe (CA24101) supported by COST (European Cooperation in Science and Technology).

\bibliographystyle{elsarticle-num}

\bibliography{main2}
    
	
\end{document}